\documentclass{article}
\usepackage[authoryear]{natbib}
\usepackage[utf8]{inputenc}
\usepackage{geometry}
\usepackage{float}
\usepackage{bm}
\usepackage{amsmath}
\usepackage{graphicx}
\usepackage{aecompl}
\usepackage{setspace}
\usepackage{lineno}
\usepackage{amsmath}
\usepackage{amssymb}
\usepackage{authblk} 
\usepackage{bbm}
\usepackage{stmaryrd}

\usepackage[colorinlistoftodos]{todonotes}
\usepackage[colorlinks=true, allcolors=blue]{hyperref} 

\title{Inferring species interactions using Granger causality and convergent cross mapping}

\author[1,2,*]{Frédéric Barraquand} 
\author[1,2]{Coralie Picoche}
\author[3]{Matteo Detto}
\author[4]{Florian Hartig}

\affil[1]{Institute of Mathematics of Bordeaux, CNRS \& University of Bordeaux, Talence, France}
\affil[2]{Integrative and Theoretical Ecology, LabEx COTE, University of Bordeaux, Pessac, France}
\affil[3]{Department of Ecology and Evolutionary Biology, Princeton University, Princeton, USA}
\affil[4]{Theoretical Ecology, University of Regensburg, Regensburg, Germany}

\date{}

\begin{document}

\maketitle
\thispagestyle{empty}

\begin{abstract}

Identifying directed interactions between species from time series of their population densities has many uses in ecology. This key statistical task is equivalent to causal time series inference, which connects to the Granger causality (GC) concept: $x$ causes $y$ if $x$ improves the prediction of $y$ in a dynamic model. However, the entangled nature of nonlinear ecological systems has led to question the appropriateness of Granger causality, especially in its classical linear Multivariate AutoRegressive (MAR) model form. Convergent cross mapping (CCM), a nonparametric method developed for deterministic dynamical systems, has been suggested as an alternative. Here, we show that linear GC and CCM are able to uncover interactions with surprisingly similar performance, for predator-prey cycles, 2-species deterministic (chaotic) or stochastic competition, as well as 10- and 20-species interaction networks. We found no correspondence between the degree of nonlinearity of the dynamics and which method performs best. Our results therefore imply that Granger causality, even in its linear MAR($p$) formulation, is a valid method for inferring interactions in nonlinear ecological networks; using GC or CCM (or both) can instead be decided based on the aims and specifics of the analysis. 

\end{abstract}
~ 
\\~\\~\\
* Corresponding author:  \texttt{frederic.barraquand@u-bordeaux.fr}\\

Published in \textit{Theoretical Ecology}, DOI: 10.1007/s12080-020-00482-7

\clearpage

\doublespacing


\section*{Introduction}

Inferring links between different species' population dynamics is a statistical endeavour with profound implications for understanding coexistence mechanisms  \citep{adler_coexistence_2010,adler_competition_2018}, food web structure and functioning \citep{berlow2004interaction,wootton2005mis}, as well as management and conservation at the ecosystem level \citep{link2002does,pikitch2004ecosystem}. However, statistically detecting such dependencies using correlative approaches can be extremely challenging \citep{coenen2019limitations,carr2019correlation}. Outside of the usual limitations induced by sample sizes, spatial or temporal co-occurrence \citep{cazelles2016theory} or co-abundance patterns \citep{stone1991conditions,loreau2008species} do not directly indicate interactions between species \citep{dormann2018biotic,blanchet2020cooccurrence}. For instance, strongly competitive communities usually show a large amount of positive associations between species abundances, not only because abiotic forcing makes synchrony the general rule \citep{loreau2008species}, but also because indirect interactions make the enemy of my enemy a friend \citep{stone1991conditions}. 
To infer dependencies between species' population dynamics, it is therefore often useful to build on a dynamic ecological and statistical theory.

A first step towards inferring interactions between species is to agree on a definition of an interaction \citep{berlow2004interaction}. For the purpose of this paper, two species $i$ and $j$ are deemed to interact if species $i$'s population growth rate is affected by the population density of species $j$ or vice-versa. This definition maps well to theoretical ecology, where communities are modelled as variations of the generalized Lotka-Volterra equations (e.g., \citealp{may1973stability,yodzis1998local,coyte2015ecology};
eq. \ref{eq:GLV}):

\begin{equation}
\frac{dN_{i}}{dt}=r_{i}N_{i}+\sum_{j=1}^{S}{g_{ij}(N_{i},N_{j})N_{j}}\label{eq:GLV}
\end{equation}
Interestingly, this definition also matches with that of statistical time series models \citep{ives2003ecs,mutshinda2009drives,mutshinda2011multispecies,hampton2013quantifying}. Embracing that ecological systems are inevitably stochastic, an interaction can therefore be defined as a link from species $j$'s density to species $i$'s per capita growth rate in a stochastic dynamical system. This has also been referred to as local dependence \citep{schweder1970composable}, dynamic causation \citep{aalen1987dynamic,aalen2012causality,sugihara2012detecting}, and Granger-Wiener causality \citep{Granger1969,geweke1982measurement,detto2012causality} in the statistics and theoretical ecology literatures.

To infer such dynamic causation from multiple time series, ecologists have used a range of statistical models, most notably Multivariate AutoRegressive models of order one, or MAR(1) models (also called VAR(1) -- vector autoregressive models -- in the econometrics
and neuroscience literatures). When the state variables are species densities, these are multispecies generalisations of the discrete-time Gompertz population growth \citep{ives2003ecs,mutshinda2009drives}, including an interaction coefficient for each species pair.
MAR($p$) models, with a maximum time lag of order $p\geq1$, generalise the MAR(1) framework to more complex dependencies over time, and have been shown to map more exactly to the celebrated 
Granger-Wiener causality concept \citep{Granger1969,sims1980,Ding2006,Chen2006,barnett2009granger,detto2012causality,barnett_mvgc_2014}. Granger-Wiener causality (usually referred to as Granger causality or GC for short) is strongly tied to the
physical notion that the cause must precede the effect. Using the temporal order of events for inferring the direction of causality matches the intuition of many biologists \citep{mayr1961cause}
and especially ecologists, familiar with predators lagging behind their prey population dynamics \citep{may1973stability}.
Granger causality combines this idea of temporal precedence of the cause with statistical prediction. If a dynamical model for time series $y$ has its in-sample predictive ability of future $y$ values improved by inclusion of time series $x$ in the predictors, we say that $x$ Granger-causes $y$. This is a purely operational definition of causality, yet it is rather general and does not specify any particular model framework. It can in principle be applied to phenomenological and mechanistic frameworks alike, and it can be extended to nonparametric and spectral settings \citep{detto2012causality}. However, the simpler parametric and linear MAR($p$) models are often preferred for Granger causality testing \citep{Lutkepohl2005}. Ecologists have in fact been using the Granger causality concept implicitly many times in the form of MAR(1) models \citep[reviewed in][]{hampton2013quantifying}.

In the last decade, new methods such as Convergent Cross Mapping (CCM; \citealp{sugihara2012detecting}), using nonlinear dynamical systems theory and attractor reconstruction, have been introduced to infer interactions between species. \citet{sugihara2012detecting} criticized the application of Granger causality concept to nonlinear dynamical systems. They deemed GC best suited for linear systems dominated by stochasticity, and unfit to model systems with a highly nonlinear (chaotic) deterministic skeleton, as according to Takens' theorem, lagged values of each variable (e.g., species density) contain information about all other linked variables in the dynamical system. This viewpoint has been subsequently adopted by many ecology studies using CCM \citep[e.g.,][]{ye2015distinguishing,ye_information_2016,deyle2016tracking,monster2017causal,harford2017non,grziwotz2018empirical}. However, while it is correct that the information contained in deterministic dynamical systems cannot be ascribed to a single component of the dynamical system (referred to as ``nonseparability'' by \citealp{sugihara2012detecting}, which is actually true for nonlinear and linear dynamical systems alike, \citealp{Granger1969,runge2014detecting}), the addition of process noise may in fact allow separating the predictive abilities of $y$ vs $(x,y)$ \citep[][p. 19]{runge2014detecting}. Given that both nonlinearity and process noise are ubiquitous in ecology, Granger causality could be a potent approach to infer interactions. Moreover, although it can seem intuitive that the linear MAR($p$) implementation of Granger causality will have difficulties with nonlinear time series, it should be noted that MAR($p$) models are usually applied to log(population sizes) in ecological settings \citep{ives2003ecs}. This log-linear scale (i) allows to transform the log-normal distribution of abundance usually found in data into a normal one and more importantly (ii) transforms multiplicative growth processes into an additive model structure. MAR($p$) models on the log-scale are therefore essentially power-law models when transformed back into the original scale, a flexible way to model monotonic nonlinearities, which is used to approximate nonlinear dynamical systems \citep{ives2003ecs}. 

Modelling has indeed shown that even the simplest MAR(1) models can be surprisingly robust to nonlinearities \citep{ives1995predicting,certain2018how}, correctly inferring the sign of interactions in the case of stochastic nonlinear competition with a fixed point and multiple predator-prey systems, including limit cycles. Further evidence that linear GC can be robust to nonlinearities comes from studies that used linear GC, nonlinear GC, and CCM, and found consistent causal answers with all three \citep{hannisdal2017common,hannisdal2018causality}. These studies provide hints that Granger causality, even in its linear MAR($p$) formulation, may apply well to stochastic and nonlinear ecological dynamical systems.

MAR($p$) models have obvious strengths for causal inference:  confidence intervals for coefficients, model selection, and other inferential tools are well understood \citep{Lutkepohl2005} based on decades of development in time series analysis. By contrast, CCM is relatively new and does not benefit (yet) from the same support from statistical and probability theory. Given the advantages stemming from the great conceptual and practical simplicity of MAR($p$) models, there is a need to better understand in which ecological scenarios linear GC can be a good approximation for interaction inference, and in which cases more sophisticated techniques are needed, such as CCM \citep{sugihara2012detecting} or entropy-based methods \citep[e.g.,][]{amblard2013relation,hannisdal2018causality}. With new monitoring tools like metabarcoding making community time series increasingly available, GC methods may become even more interesting for ecologists. For example, Granger-causality techniques are currently gaining traction in the rapidly evolving microbiome field that attempts at inferring interactions from metabarcoding data on microorganisms \citep{gibbons2017two,mainali2019detecting,carr2019correlation}.

In this article, we evaluate the performance of linear MAR($p$) models and compare it to CCM on a number of ecological examples for which CCM is currently thought to be more appropriate. We demonstrate that criticism of the Granger causality concept by \citet{sugihara2012detecting} may have been induced by nonstandard model selection and evaluation techniques. Using simpler model selection techniques, routinely used
by statisticians to infer the lag order $p$ of MAR($p$) models as well as their parametric structure \citep{Lutkepohl2005}, we show that Granger causality techniques can infer interactions in nonlinear time series surprisingly well. Granger causality and CCM either both work well or both fail to some degree for most case studies, which suggests that seemingly different causality concepts might in fact share hidden similarities. Throughout our analysis, we take care to consider both statistical significance and effect sizes of causal inferences. We then demonstrate that MAR($p$) modelling can be scaled up to large interaction networks using either appropriate model regularization techniques (based on a structured version of the LASSO) or pairwise inference with an appropriate false discovery rate correction. A comparison to CCM is provided in the latter case. 

\section*{Methods and models}

To start, we recall the basics of Granger causality concepts and MAR($p$) modelling - Multivariate AutoRegressive modelling of order $p$ - which is the most common way to assess Granger causality (though by no means the only one, see e.g. \citealp{detto2012causality} for a nonparametric and spectral Granger approach, \citealp{barnett_mvgc_2014} for parametric and spectral approaches). We describe shortly thereafter convergent cross mapping \citep{sugihara2012detecting}, which takes a different approach to causal inference, based on dynamical systems theory and state-space reconstruction. We then describe the real datasets and numerical simulations that will be used for evaluating causal inference methods.

\subsection*{Causality concepts}

\subsubsection*{Granger causality and MAR($p$) implementation}

Formally, time series $\mathbf{x}=(x_{t})_{t\in \llbracket 1,T \rrbracket}$ Granger-causes time series $\mathbf{y}=(y_{t})_{t\in \llbracket 1,T \rrbracket}\Leftrightarrow$
including $x$ in a time series model for $y$ improves in-sample
prediction of $y$. In the MAR($p$) framework, this translates into
performing two autoregressive model fits to explain time series
$\mathbf{y}$, one with only $y$ values and one with both $y$ and $x$ values:

\begin{align}
y_{t}=\sum_{i=1}^{p} \alpha_{i}y_{t-i}+\eta_{t}, \; \eta_t \sim \mathcal{N}(0,\sigma_\eta^2) \label{eq:directGC}\\
y_{t}=\sum_{i=1}^{p}\beta_{1i}x_{t-i}+\sum_{i=1}^{p}\beta_{2i}y_{t-i}+\epsilon_{t}, \; \epsilon_t \sim \mathcal{N}(0,\sigma_\epsilon^2)\label{eq:GC_bis}.
\end{align}

Granger causality is inferred if $\sigma_{\epsilon}^{2}<\sigma_{\eta}^{2}$ and such difference is statistically significant. A simple measure of effect size is therefore the log ratio of the sum of squared residuals $G_{x \rightarrow y} = \ln \left(\frac{\sigma_{\eta}^{2}}{\sigma_{\epsilon}^{2}}\right)$ \citep{geweke1982measurement,detto2012causality}. 
When more than two variables are considered, pairwise GC has to be differentiated
from conditional GC \citep{geweke1984measures,barnett_mvgc_2014}. Conditional GC occurs whenever a third variable
$z$ is considered and corrected for. When fitting a MAR($p$) model to more than two species, we would typically be interested in conditional GC rather than pairwise GC, with conditional GC correcting for the effects of non-focal species and abiotic covariates.
For instance, let us consider a MAR(1) model (eq. \ref{eq:MAR1})
with 3 species in the classic form of \citet{hampton2013quantifying}, where $\mathbf{N}_{t}$ is the vector of population densities:

\begin{equation}
\mathbf{x}_{t}=\ln(\mathbf{N}_{t}),\,\,\,\mathbf{x}_{t+1}=\mathbf{a}+\mathbf{Bx}_{t}+\mathbf{Cu}_{t}+\mathbf{e}_{t},\mathbf{e}_{t}\sim\mathcal{N}_{3}(\mathbf{0},\bm{\Sigma})\label{eq:MAR1}.
\end{equation}

Here, the entries in the interaction matrix $\mathbf{B}$ are defined by
\begin{center}
\begin{equation}
\mathbf{B}=\begin{pmatrix}b_{11} & b_{12} & b_{13}\\
b_{21} & b_{22} & b_{23}\\
b_{31} & b_{32} & b_{33}
\end{pmatrix}\label{eq:B_mAR3}
\end{equation}
\par\end{center}

and $\mathbf{C}$ is a matrix representing the effect of environmental covariates $\mathbf{u}_t$ \citep{ives2003ecs,hampton2013quantifying}. Whenever $b_{12}$ is significantly different from zero, we
have a causal influence $x_{2}\rightarrow x_{1}|(x_{3},\mathbf{u})$,
that is, an influence of $x_{2}$ on $x_{1}$ conditional to the population
density $x_{3}$ of species 3 and all the control environmental variables
in the vector $\mathbf{u}$. 

Using centered data so that the intercept
disappears, the MAR($p$) model is defined as

\begin{equation}
\mathbf{y}_{t+1}=\sum_{q=1}^{p}\mathbf{B}^{(q)}\mathbf{y}_{t-q+1}+\mathbf{e}_{t},\;\mathbf{e}_{t}\sim\mathcal{N}_{d}(\mathbf{0},\bm{\Sigma})\label{eq:MARp}
\end{equation}

where $d$ is the number of system components (individual time series). For a general definition of causal effects, we dropped $\mathbf{u}_{t}$ from eq. \ref{eq:MARp}, as it corresponds to a special case where a subset $\mathbf{u}_{t}$ of the variables $\mathbf{y}_{t}=(\mathbf{x}_{t},\mathbf{u}_{t})'$ has a one-way causal impact (i.e., $\mathbf{u}_{t}$ affects $\mathbf{x}_{t+1}$ but not the other way around, which can be specified as well by forcing the $\mathbf{B}^{(q)}$ matrices to contain some zeroes). The condition for an interaction from
system component $j$ to system component $i$ given all other system components (either species densities or environmental variables) then becomes, in a general MAR($p$) setting (according to
eq. \ref{eq:MARp}): 
\begin{equation}
\exists b_{ij}^{(q)}\neq0\Leftrightarrow y_{j}\rightarrow y_{i}|(y_{1},...,y_{j-1},y_{j+1},...,y_{d})\label{eq:conditionGC_MARp}
\end{equation}
where each time lag is indexed by $q$. 
Conversely, pairwise GC testing between $y_{i}$ and $y_{j}$ is assessed through a bivariate autoregressive model for each $(i,j)$ pair, and therefore uses a considerably lower-dimensional model, although it may require a false discovery correction to attain meaningful statistical significance (see next section).

To implement these concepts in practice, we fitted MAR($p$) models using the package \verb|vars| in \verb|R| (version 3.4.4), which uses ordinary least squares for estimation. We mainly used the BIC as a default for lag order selection, although we also considered other information criteria (see below). The presence of Granger causality was assessed by the statistical significance and magnitude of the interaction matrix coefficients, and more directly using parametric significance tests for nested models. For pairwise Granger causality testing, we used the function \verb|grangertest| in the R package \verb|lmtest| \citep[][v0.9-36]{zeileis2002diagnostic} which performs a Wald test for nested models (based on the statistical significance of MAR model coefficients). For conditional Granger causality testing, we used the function \verb|causality| in package \verb|vars| \citep[][v1.5-3]{pfaff2008var} which provides F-tests for nested models. Both tests and implementations provided similar answers when compared.

\subsubsection*{Granger causality in high-dimensional models}

If we have a large number of time series, corresponding to many species, fitting full MAR($p$) models (i.e., models that account for all possible interactions without additional constraints) becomes impractical, unless those time series are extremely long \citep{michailidis2013autoregressive}.
For $d$ species and $p$ time lags, a $d\times d\times p$
dimensional model needs to be fitted to the data. For instance, 10 species with $p=2$ yields $2\times10\times10=200$ parameters in the interaction
\textbf{B}$^{(q)}$ matrices only. Even a simpler MAR(1) model would be impossible to fit properly
without a set of time series of length above 100 (or some added regularization). Preliminary simulations \citep{certain2018how} suggest that a nonlinear, stochastic ecological system of dimension 10 or 12 requires approximately time series of length 500 to 800 to be fitted properly without implementing additional constraints. To deal with high-dimensionality, we considered two solutions:
\begin{itemize}
\item Pairwise Granger causality testing with False Discovery Rate (FDR)
correction (Benjamini-Hochberg), with a philosophy similar to \citet{mukhopadhyay2006causality}.
This is done by fitting bivariate MAR($p$) models, testing for Granger causality
in both directions, and then re-adjusting the p-values obtained through
the \citet{benjaminihochberg1995} correction.
\item LASSO-penalized MAR(1) models with structured penalties, using the
R package \verb|SIMoNe|\textbf{ }\citep[][v1.0-3]{chiquet2008simone,charbonnier2010weighted}.
This allows to estimate (through non-zero interaction coefficients) conditional Granger causality. A naive idea would be to use the classic LASSO (Least Absolute Shrinkage and Selection
Operator, \citealp{tibshirani2015statistical}) to set some of the
coefficients to zero. Unfortunately, this approach is known to yield
substantial bias whenever there is an important structure (here, modular) in the network \citep{charbonnier2010weighted}. The technique that we used explicitly accounts for network structure in addition to selecting coefficients with the LASSO, and is described in Electronic Supplementary Material Appendix~\ref{sec:LASSO-VAR}.
\end{itemize}

\subsubsection*{Convergent cross mapping}

Convergent cross mapping (CCM) was proposed by \citet{sugihara2012detecting} as an alternative  nonparametric method to detect dependencies between time series. CCM relies on state-space reconstruction. We assume two time series $\mathbf{x}=(x_{t})_{t\in\llbracket 1,T \rrbracket}$ and $\mathbf{y}=(y_{t})_{t\in\llbracket 1,T \rrbracket}$ as previously. The attractor
manifold $M_{X}$ is constructed as a set of \emph{E}-dimensional
vectors $\tilde{\mathbf{x}}(t)=(x_t,x_{t-\tau},x_{t-2\tau},...,x_{t-(E-1)\tau})$
for $t=1+(E-1)\tau$ to $t=T$. $E$ is the embedding dimension, denoting
how many time lags one counts back in time. This set of vectors $\{ \tilde{\mathbf{x}}(t)\}$ constitutes
the reconstructed manifold. We now find the $E+1$ nearest neighbours
of each $\tilde{\mathbf{x}}(t)$ in $M_{X}$. Their time indices are denoted
$t_{1},...,t_{E+1}$. The reconstruction of $y_{t}$ from $M_{X}$
proceeds as follows:

\begin{equation}
 \hat{y}(t)|M_{X}=\sum_{i=1}^{E+1}w_{i}y(t_{i})
\end{equation}

with $w_{i}=~u_{i}/\sum_{j=1}^{E+1}u_{j}$, and $u_{j}=\exp\left(\frac{-d(\tilde{\mathbf{x}}(t),\tilde{\mathbf{x}}(t_{j}))}{d(\tilde{\mathbf{x}}(t),\tilde{\mathbf{x}}(t_{1}))}\right)$
where $d(\tilde{\mathbf{x}}(t),\tilde{\mathbf{x}}(t_{1}))$ is the
minimal distance between $\tilde{\mathbf{x}}(t)$ and all other embedded points.

The cross-map skill from $X$ to $Y$ is then measured by the correlation
coefficient $\rho(\mathbf{y},\hat{\mathbf{y}}|M_{X})>0$, which increases
with the size $L$ of the library of points used to reconstruct the manifold $M_{X}$ if $Y$ causes $X$. The surprising thing here is
that predicting $Y$ by $M_{X}$ is equivalent to $Y$ causing
$X$ and not the other way around \citep{sugihara2012detecting}.
Hence, to know if $X$ causes $Y$, we look at $\rho(\mathbf{x},\hat{\mathbf{x}}|M_{Y})$.

Due to the absence of a parametric model, there is no formula for the p-value related to the CCM skill $\rho$. Several p-value formulations have been proposed under the null hypothesis of no causality from $X$ to $Y$:
\begin{itemize}
\item \citet{cobey2016limits} suggested $p(X\nrightarrow Y)=\frac{1}{n}\sum_{i=1}^{n}\mathbbm{1}_{i}\left(\rho(\mathbf{x}_{\text{b},i},\hat{\mathbf{x}}_{\text{b},i}|M_{Y,\text{Lmax}})<\rho(\mathbf{x}_{\text{b},i},\hat{\mathbf{x}}_{\text{b},i}|M_{Y,\text{Lmin}})\right)$
where $n$ is the number of libraries of size $L$ that were used
to build $M_{Y}$ and $\mathbf{x}_{\text{b},i}$ are resampled (bootstrapped) values of the vector $\mathbf{x}$. $M_{Y,\text{Lmax}}$ (respectively, $M_{Y,\text{Lmin}}$)
is the manifold constructed with the maximum (respectively, minimum)
library size. Two versions of this p-value can be computed depending on whether one samples with replacement (the bootstrap) for the libraries or without replacement (in which case $M_{Y,\text{Lmin}}$ varies but not $M_{Y,\text{Lmax}}$).
\item When two species are forced by a shared forcing driver (e.g., seasonal temperature), spurious causality can emerge. This can be corrected by computing $n$ surrogate time series $\mathbf{x}_{\text{surr},i}$, that keep the periodicity of the signal but shuffle its residuals, so that cross-correlations containing causal information
are ``erased''. Cross-mapping is then computed on the surrogates and compared to the real value \citep{deyle2016global}. In this case,
$p(X\nrightarrow Y)=\frac{1}{n}\sum_{i=1}^{n} \mathbbm{1}_{i}\left(\rho(\mathbf{x}_{\text{real}},\mathbf{\hat{x}}_{\text{real}}|M_{Y})<\rho(\mathbf{x}_{\text{surr},i},\mathbf{\hat{x}}_{\text{surr},i}|M_{Y})\right)$. In fact, the more exact formula is $p(X\nrightarrow Y)=\frac{r+1}{n+1}$ where $r=\sum_{i=1}^{n} \mathbbm{1}_{i}\left(\rho(\mathbf{x}_{\text{real}},\mathbf{\hat{x}}_{\text{real}}|M_{Y})<\rho(\mathbf{x}_{\text{surr},i},\mathbf{\hat{x}}_{\text{surr},i}|M_{Y})\right)$, following \citet{north2002note}, which we employed here. 
\item Given that surrogate-based p-values, required in the shared abiotic driver case (item above), were found to perform better than alternative p-values in most contexts, we computed those systematically for all simulations. For model simulations where there was no confounding abiotic driver, surrogates were only computed by permutation of the time series. This was found to be simpler and more efficient than other techniques to provide statistical significance for CCM (Appendix~\ref{subsec:Choice-of-p-values}).
\end{itemize}

The analyses have been performed using the package \verb|rEDM| \citep[][v0.7.1]{ye2018redm}. For each time series, we retrieved the best embedding dimension (which maximizes the forecast skill of the simplex, \citealp{sugihara1990nonlinear}) and used it in the cross-mapping function, with 100 different libraries for each library size and maximum library size depending on the length of the time series (300 timesteps if not mentioned otherwise). The libraries were obtained with random draws without replacement from the original time series. In high-dimensional cases, we used the same Benjamini-Hochberg correction as for GC. 

\subsubsection*{Evaluating GC and CCM}

GC was evaluated using a criterion of p-value below 0.1 (0.2 in a high-dimensional setting) or an effect size $G_{x \rightarrow y}>0.04$ (threshold found and justified in Appendix~\ref{subsec:Choice-of-p-values}), or both criteria simultaneously. Our philosophy here has been to evaluate causality based on both statistical significance and effect sizes, to avoid the well-known drawbacks of considering solely statistical significance. The same philosophy is used for evaluating CCM, where we considered two thresholds for correlation coefficient $\rho$ of the cross-mapping (measuring the effect size), 0.1 and 0.2. These thresholds also originate from preliminary analyses presented in Appendix~\ref{subsec:Choice-of-p-values}. 

For each case study, we compared the values of classical scores such as recall or sensitivity (fraction of true interactions $TP$ found over the total number of true interactions, $\frac{TP}{TP+FN}$, where $FN$ are false negatives) and the specificity (fraction of true negatives $TN$ over the total number of negatives, $\frac{TN}{TN+FP}$, where $FP$ are false positives).
Additionally, we measured similarity between GC- and CCM-detected causalities at the level of individual time series -- within a single parameter set and model -- to see if they detect matching causalities or have some degree of complementarity. We used the Sokal Michener index $I_{SM}=\frac{\sum \mathbbm{1}_{11}+\sum \mathbbm{1}_{00} }{\sum \mathbbm{1}_{11}+\sum \mathbbm{1}_{10}+\sum \mathbbm{1}_{01}+\sum \mathbbm{1}_{00}}$ where $\sum \mathbbm{1}_{11}$ indicates the number of simulations for which GC and CCM both indicate causality, $\sum \mathbbm{1}_{10}$ the number of simulations for which GC indicates causality and CCM does not, etc. 

\subsection*{Simulated and real case studies of interacting species population dynamics}

\subsubsection*{Real data: Veilleux's predator-prey cycles}

\begin{figure}[H]
\begin{centering}
\includegraphics[width=0.95\textwidth]{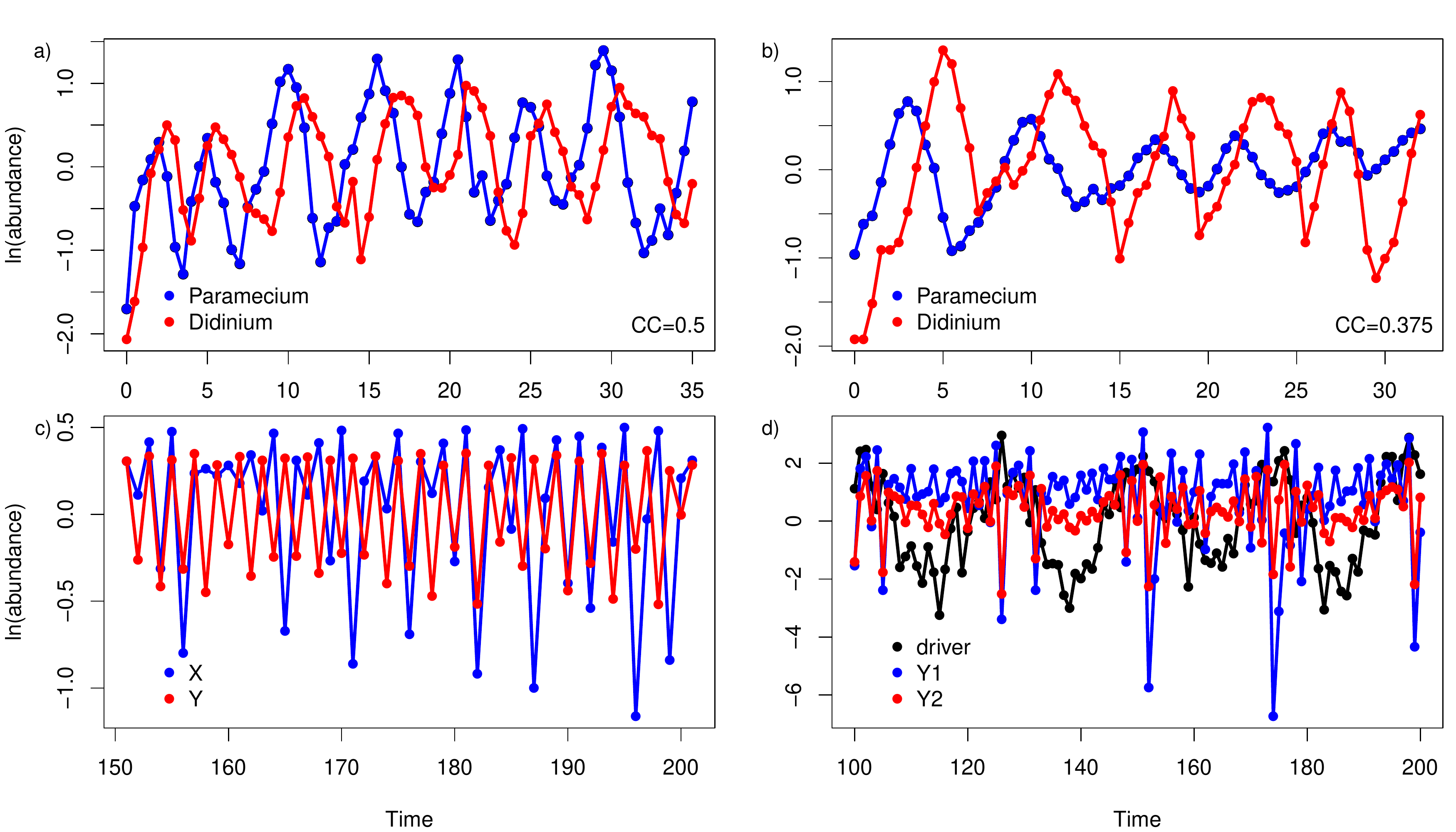}
\par\end{centering}
\caption{Time series of small-community models. Veilleux's predator-prey data
are shown in (a) (dataset CC05) and (b) (dataset CC0375); an example
simulation for the 2-species chaotic model is shown in panel (c) and
a simulation of the competition model including an environmental driver
is illustrated in panel (d).\label{fig:TS}}
\end{figure}

The first two datasets that we consider are taken from \citet{veilleux1979analysis}
and have been analysed by other authors with mechanistic models that demonstrated two-way coupling \citep{jost2000testing}, plausibly with limit cycle behaviour (Fig.~\ref{fig:TS}a,b). We additionally created 500 simulated time series from MAR($p$) models that best fitted to this dataset, to provide a `linear' dynamical version of this empirical system. 

\subsubsection*{Deterministic chaos in two-species competition models}

Our second case study is the two-species discrete-time logistic competition model (Fig.~\ref{fig:TS}c) used in \citet{sugihara2012detecting} to evaluate the performance of CCM: 

\begin{align}
x_{t+1} & =x_{t}(3.8-3.8x_{t}-0.02y_{t})\label{eq:two_species_chaotic_compet}\\
y_{t+1} & =y_{t}(3.5-3.5y_{t}-0.1x_{t})
\end{align}

Model parameters are identical to \citet{sugihara2012detecting}, which places this model in the chaotic regime (Lyapunov exponent LE $= + 0.41$). This case study therefore constitutes a strong test of the log-linear MAR($p$) framework. The only setting that was modified compared to \citet{sugihara2012detecting} is the initial condition, which was randomly drawn from a Uniform(0,1) distribution 500 times. Although we acknowledge that ``mirage correlations''
can occur in some datasets, we aimed at reproducing the full distribution
of what this model can provide, as there are no justifications to
favour one specific set of initial conditions (outside of illustration
purposes). The sample size is taken to be $T=300$ as in \citet{sugihara2012detecting},
after an initial run of 500 time steps that are discarded to remove transients.

Because a method that finds no interactions whenever absent (i.e., no false positives) is as important as one that finds interactions whenever they are present, we additionally created simulations without interactions:

\begin{align}
x_{t+1} & =x_{t}(3.8-3.8x_{t}-0\times y_{t})\label{eq:two_species_chaotic_compet-1}\\
y_{t+1} & =y_{t}(3.5-3.5y_{t}-0\times x_{t})
\end{align}

We evaluated both GC and CCM's ability to find no interactions
between these time series.

\subsubsection*{Two-species stochastic and nonlinear dynamics, including environmental
drivers}

We consider here a stochastic two-species competition model, with Lotka-Volterra interactions in discrete time and a Ricker type of multispecies density-dependence:

\begin{align}
N_{1,t+1} & =N_{1,t}\exp(3-4N_{1,t}-2N_{2,t}+\epsilon_{1,t})\label{eq:twoSpeRickerStoch_compet}\\
N_{2,t+1} & =N_{2,t}\exp(2.1-0.31N_{1,t}-3.1N_{2,t}+\epsilon_{2,t}\label{eq:twoSpeRickerStoch_compet_2}).
\end{align}

An illustration is provided in ESM~ \ref{sec:2sppRickerillustration}. This case was already investigated in \citet{certain2018how}, including as well an environmental driver on species 1 (but not species 2). The model of eqs.~\ref{eq:twoSpeRickerStoch_compet}--\ref{eq:twoSpeRickerStoch_compet_2} has a stochastic Lyapunov exponent (SLE) of $-0.18$, and therefore is not sensitive to initial conditions when perturbed by noise \citep{ellner2005can}. The stochastic Lyapunov exponent, as elsewhere in this manuscript, was computed following \citet{dennis2001estimating}. The dynamics of the corresponding deterministic skeleton are a two-cycle when there are interactions between species, while without interactions, species 1 is chaotic and species 2 has a two-cycle. 

As a fourth case study, we created a scenario to investigate the effect of environmental drivers on the estimation of species interactions (Fig.~\ref{fig:TS}d). This is done with a variant of eqs.~\ref{eq:twoSpeRickerStoch_compet}--\ref{eq:twoSpeRickerStoch_compet_2} by adding an environmental driver $u_{t}$ that
has the same effect on both species, which constitutes a challenge
for any causal method ($u_{t}$ is a confounding variable):

\begin{align}
N_{1,t+1} & =N_{1,t}\exp(3+0.5u_{t}-4N_{1,t}-2N_{2,t}+\epsilon_{1,t})\label{eq:StochTwoSpecies_withSynchroDriver}\\
N_{2,t+1} & =N_{2,t}\exp(2.1+0.5u_{t}-0.31N_{1,t}-3.1N_{2,t}+\epsilon_{2,t}).
\end{align}

We considered, as in the deterministic case, the counterparts of the
above models where the interspecific interactions are set to zero,
i.e.,

\begin{align}
N_{1,t+1} & =N_{1,t}\exp(3+0.5u_{t}-4N_{1,t}-0\times N_{2,t}+\epsilon_{1,t})\label{eq:twoSpeciesStoch_wDriver_noInteractions}\\
N_{2,t+1} & =N_{2,t}\exp(2.1+0.5u_{t}-0\times N_{1,t}-3.1N_{2,t}+\epsilon_{2,t}).
\end{align}

We ran 500 simulations for each model. The noise was set so that $\epsilon_{i,t}\sim\mathcal{N}(0,\sigma^{2})$
i.i.d. with $\sigma^{2}=0.01$, but one should keep in mind that the addition of the environmental drivers increases the level of noise in this system. This slightly increases the Lyapunov exponent ($\approx -0.15$ for the coupled system).

\subsubsection*{Ten- and twenty-species interaction webs}

We simulated a ten-species model, which generalises the two-species Ricker competition to more species and more interaction types, with added stochasticity ($\sigma^{2}=0.1$). This model therefore represents a considerable challenge to interaction inference, due to the large quantity of potential false positives (many zero interactions) combined to both nonlinear dynamics and stronger stochasticity. The dynamical equation can be written as

\begin{equation}
\mathbf{N}_{t+1}=\mathbf{N}_{t}\circ\exp(\mathbf{r}+\mathbf{A}\mathbf{N}_{t}+\mathbf{e}_{t}),\mathbf{e}_{t}\sim\mathcal{N}(0,\sigma^{2}\mathbf{I})\label{eq:LV10species}
\end{equation}
where $\mathbf{N}$ is the abundance vector, $\sigma^{2}=0.1$ is the process noise variance, 
and the interaction matrix $\mathbf{A}$ is defined to be
\begin{equation}
\ensuremath{\mathbf{A}=\begin{pmatrix}-4 & -2 & -0.4 & 0 & 0 & 0 & 0 & 0 & 0 & 0\\
-0.31 & -3.1 & -0.93 & 0 & 0 & 0 & 0 & 0 & 0 & 0\\
0.636 & 0.636 & -2.12 & 0 & 0 & 0 & 0 & 0 & 0 & 0\\
-0.111 & -0.111 & 0.131 & -3.8 & 0 & 0 & 0 & 0 & 0 & 0\\
0 & 0 & 0 & 0.5 & -2 & -2 & -0.4 & 0 & 0 & 0\\
0 & 0 & 0 & 0 & -0.31 & -3.1 & -0.93 & 0 & 0 & 0\\
0 & 0 & 0 & 0 & 0.636 & -0.636 & -2.12 & 0 & 0 & 0\\
0 & 0 & 0 & 0 & 0 & 0 & 0 & -4 & -2 & -0.4\\
0 & 0 & 0 & 0 & 0 & 0 & 0 & -0.31 & -3.1 & -0.93\\
0 & 0 & 0 & 0 & 0 & 0 & 0 & 0.636 & 0.636 & -2.12
\end{pmatrix}}.
\end{equation}

This Lotka-Volterra model has a stochastic Lyapunov exponent (SLE) of +0.33. This positive SLE clearly places the model in a noisy chaotic regime \citep{ellner2005can}. In addition, we used the Jacobian matrix (of the model in eq.~\ref{eq:LV10species}) as the interaction matrix of a MAR(1) model, which has
therefore comparable interaction strengths but non-chaotic dynamics. In this case, the dynamical equation is written as 
\begin{eqnarray}
    \mathbf{x}_{t+1}&=&\mathbf{J}\mathbf{x}_{t}\\
    \text{with }J_{ij}&=&\delta_{ij} + a_{ij} N_{j}^{*}
\end{eqnarray}
where $\mathbf{x}_t=\mathbf{n}_t-\mathbf{n}^*$, with $n_t=\ln(N_t)$ and $\mathbf{n}^*$ being the equilibrium on log-scale, $\delta_{ij}=1$ if $i=j$ and $\delta_{ij}=0$ otherwise (see derivation in the ESM Appendix~\ref{sec:jacobian}). By definition, such MAR(1) models have a single fixed point forced by stochasticity when stable \citep{ives2003ecs}: they cannot exhibit chaos and therefore exhibit negative SLEs. We ran 25 simulations over 500 time steps with different initial conditions, for both the chaotic LV model and its log-linear MAR(1) counterpart. We analysed the last 300 time steps. We slightly modified this model to scale it up to 20 species, with a structure that is still very modular (eq.~\ref{eq:mat_20species} in ESM~\ref{sec:20sp-matrix}). For the 20-species model, we also compared Ricker and MAR(1) dynamics for 25 different simulations over 1000 time steps and analysed the last 700. In the 20-species case, coefficients were drawn from a probability distribution (eqs. \ref{eq:coef_20species}--\ref{eq:coef_20species_condition}) and therefore differ from one simulation to the next, although we have taken care to avoid coefficients too close to zero by imposing a lower bound:
\begin{equation}
a_{ij}=\chi_{ij} \left[ a_{min}+(a_{max}-a_{min})\text{Beta}(2,2)\right] \label{eq:coef_20species}
\end{equation}
with the bounds of the interaction coefficient selected as
\begin{equation}
(a_{min},a_{max})=\begin{cases}
(0.05,0.1) & \forall i\neq j,\text{ with probability 0.2 (positive interaction)}\\
(-0.2,-0.1) & \forall i\neq j,\text{ with probability 0.8 (negative interaction)}\\
(-0.8,-0.3) & \forall i=j
\end{cases}\label{eq:coef_20species_condition}.
\end{equation}
This construction of the interaction coefficients allows to have some
realistically strong dominance of the diagonal coefficients, a certain
percentage of weak facilitation (20\%), and marked competition between species whenever interactions are allowed by the network structure.
The 20-species Ricker models thus constructed have SLEs slightly below zero (mean = -0.05, SD = 0.04), and are therefore less nonlinear (\textit{sensu} sensitivity to initial conditions) than the 10-species models considered above. The deterministic skeleton of the model has additionally always a stable fixed point. For all datasets, real and simulated alike, the data have been log-transformed and centered before analysis, which is required for GC and does not change performance for CCM (Appendix~\ref{sec:log-transfo}). We used a FDR of 20\% in all pairwise high-dimensional analyses. 

\section*{Results}

In the following, we first report both GC/MAR($p$) and CCM results for each dataset or model. The results are then summarized in Fig.~\ref{fig:diagnostics_vs_Lyapunov}. 

\subsection*{Real data: Veilleux's predator-prey cycles}

\begin{figure}[H]
\begin{centering}
\includegraphics[width=0.65\textwidth]{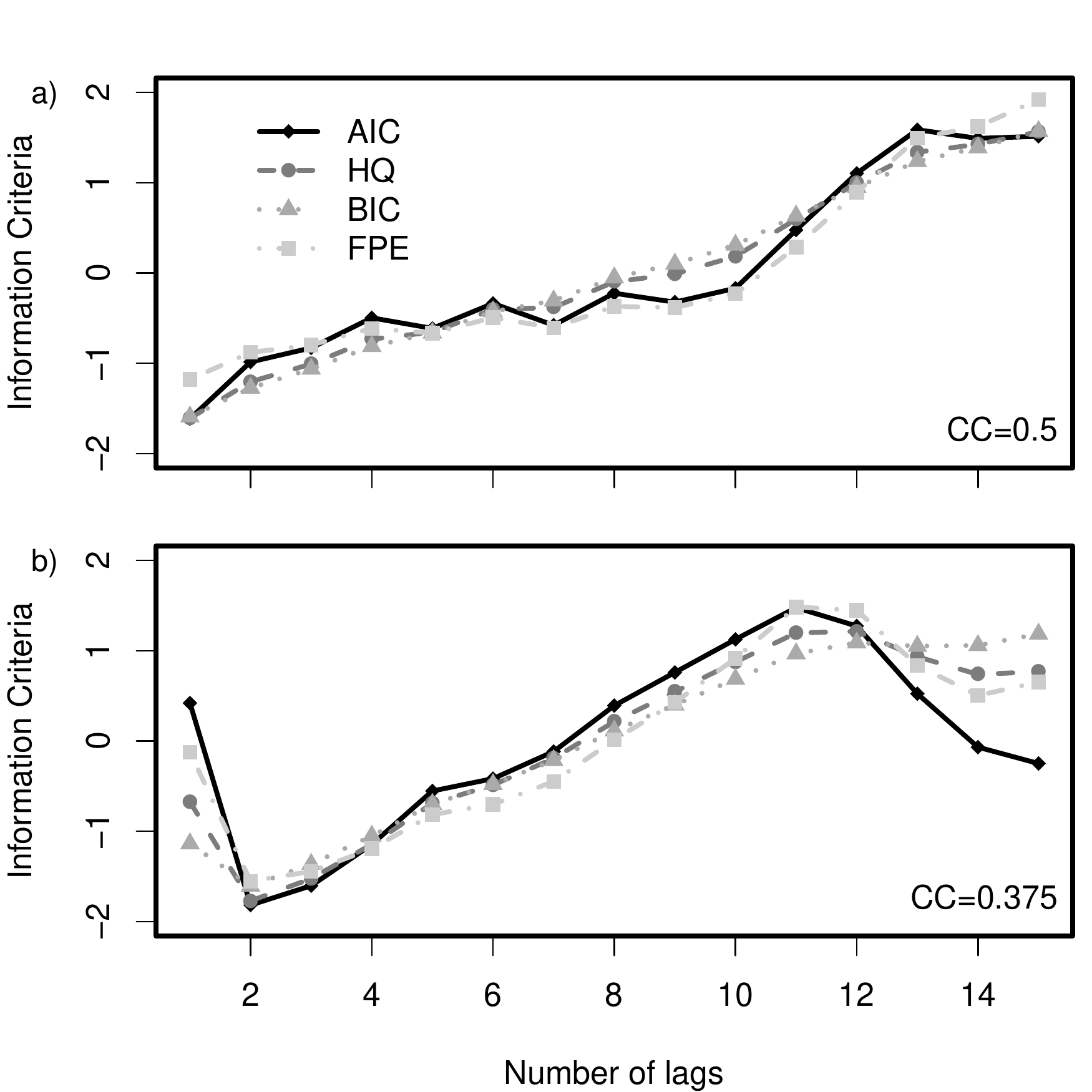}
\par\end{centering}
\caption{Model information criteria as a function of lag order for the predator-prey
data, for the two datasets. a) CC = 0.5 and b) CC = 0.375\label{fig:ResultsVeilleux_lagOrder}}
\end{figure}

On those two datasets, both GC and CCM correctly identify the two-way predator-prey coupling. Surprisingly, CCM also identifies reciprocal causal influences in the linear MAR($p$) approximation. Model selection of MAR($p$) model by all information criteria selected
a lag $p=1$ for the CC0.5 dataset and a lag of 2 for the CC0.375 dataset
(Fig. \ref{fig:ResultsVeilleux_lagOrder}). The p-values for the GC
test (null hypothesis: ``no GC'') and associated effect sizes demonstrate convincingly that the ``no GC'' hypothesis can be rejected, for both datasets (Table \ref{tab:P-valuesVeilleux}).

\begin{table}[H]
\centering \begin{tabular}{c|cc|cc}    \hline   
Dataset & CC = 0.5& & CC = 0.375&\\  \hline  
 Lag $p$ in MAR($p$) & 1 && 2& \\  \hline  
Metrics & p-val & $G_{x \rightarrow y}$ & p-val & $G_{x \rightarrow y}$\\ \hline
 1 $\rightarrow$ 2 & $2.79 \times 10^{-11}$& 0.76  & 0.0409 & 0.09\\  
 2 $\rightarrow$ 1 & $1.76 \times 10^{-14}$& 1.02 & 0.0464 & 0.10\\  \hline \end{tabular}

\caption{P-values for $H_{0}$: $\{$ No Granger causality between $x$ and $y$ $\}$ and effect sizes of GC.\label{tab:P-valuesVeilleux}}
\end{table}

\begin{figure}[H]
\begin{centering}
\includegraphics[width=0.9\textwidth]{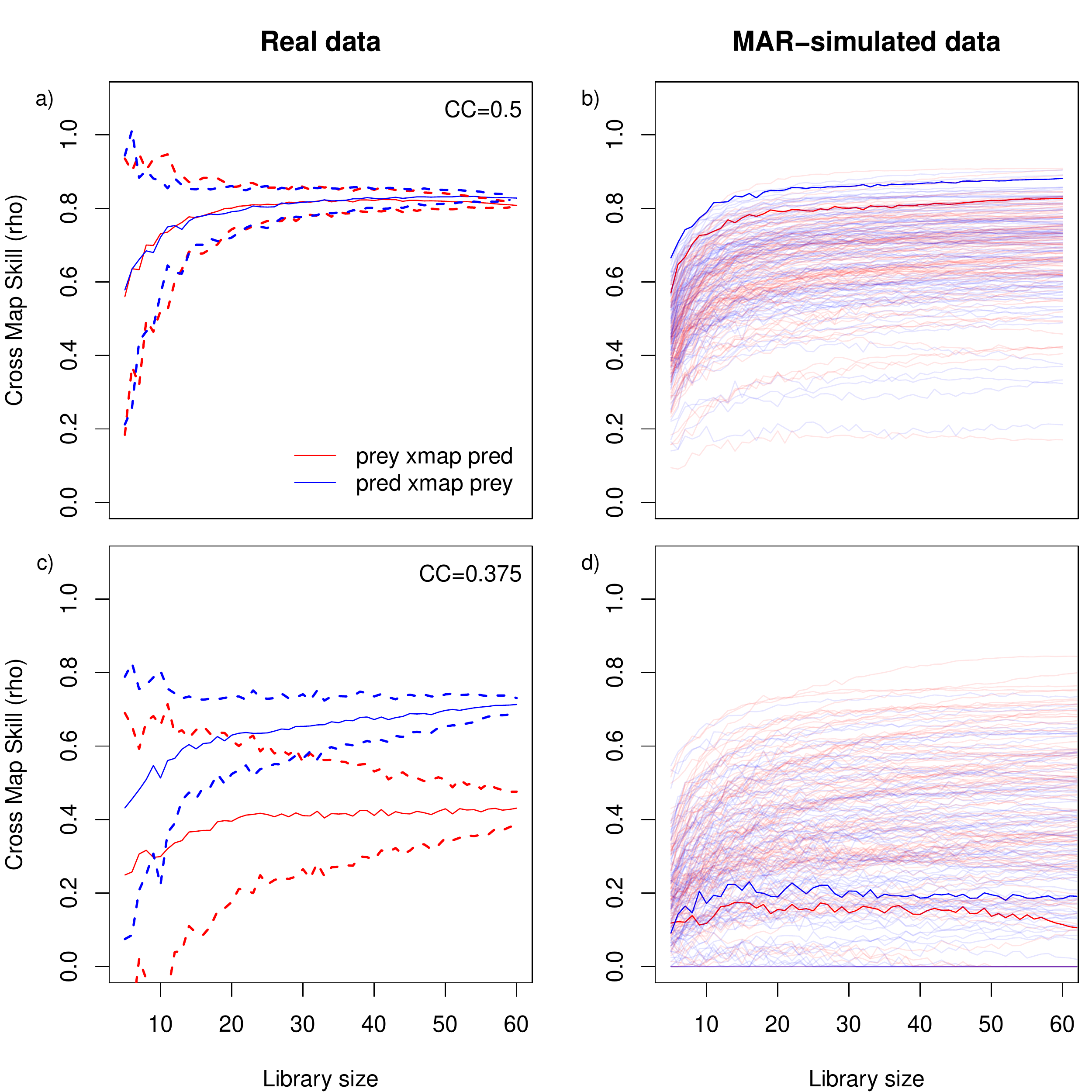}
\par\end{centering}
\caption{Convergent cross mapping for Veilleux's CC0.5 dataset (a and b) and
CC0.375 dataset (c and d). Dotted lines represent the confidence bands  (+/- 2 SD),
obtained by bootstrapping. b) and d) present CCM analyses on data that were simulated using the best-fitting MAR($p$) models to the Veilleux datasets. \label{fig:CCMVeilleux}}
\end{figure}

CCM also demonstrates bi-directional causality, as demonstrated by
the substantial increase in $\rho(\mathbf{x},\hat{\mathbf{x}}|M_{Y})$ with library size $L$ in both directions (Fig.~\ref{fig:CCMVeilleux}a and c). This is true for the real data (with or without log transformation, ESM Fig.~\ref{fig:log-transfo-veilleux}), but also many MAR(1)-simulated dataset using the fitted MAR(1) as the data-generating model (Fig.~\ref{fig:CCMVeilleux}b and d).

\subsection*{Deterministic chaos in two-species competition models}

In the two-species chaotic competition model, high-order temporal lags
tend to be selected (ESM Fig.~\ref{fig:ResultsDeterCompet_lagOrder})
despite the single time lag considered in the simulation model (i.e.,
higher nonlinearity is expressed as high-order lags). The optimal lag is $p=7$ for the model with interactions ($p=3$ without interactions), for which we report the results in Table~\ref{tab:PropCausality_DeterModel}. Despite this potential overparameterization, the GC tests show that causality is
detected for most lag orders (including $p=7$) whenever causality is present (ESM Fig.~\ref{fig:GC_deterministic}). Further, the tests are not able to reject the null hypothesis of no GC when GC is not present (Table~\ref{tab:PropCausality_DeterModel}, Fig.~\ref{fig:GC_deterministic} in Appendix for $p<7$), and the false positive rate is close to the assumed 10\%. GC performs therefore surprisingly well in this chaotic context. CCM
performs well when considering a simulation model with interactions,
but not substantially better than GC concerning the weak causal effect $2\rightarrow1$
(Table~\ref{tab:PropCausality_DeterModel}). The false positives (type I error) are somewhat higher in one causal direction for CCM (Table~\ref{tab:PropCausality_DeterModel}), up to 27\% (instead of 10\%) unless all $\rho$ values below 0.2 are discarded (i.e., thresholding based on effect sizes). This is because a large number of simulations still show an increase of $\rho$ with the library size $L$ even though there is no causality (ESM Fig. \ref{fig:CCM_Deter}). This may have been missed in \citet{sugihara2012detecting} because specific sets of initial conditions were selected, instead of drawing 500 at random as done here.

\begin{table}[H]
\begin{centering}
\begin{tabular}{c||ccc|ccccc||c}
\hline 
Method & Granger & causality &  & CCM &  &  &  & &$I_{SM}$ \tabularnewline
\hline 
Thresholds & pval\textless 0.1 & $G_{x \rightarrow y}$ \textgreater 0.04 & both & pval\textless 0.1 & $\rho$\textgreater 0.1 & $\rho$\textgreater 0.2 & both0.1 & both0.2& both\tabularnewline
\hline 
With inter. &  &  &  &  &  &  & && \tabularnewline
$1\rightarrow2$ & 100.0 & 100.0 & 100.0 & 100.0 & 100.0 & 100.0 & 100.0 & 100.0 & 1\tabularnewline
$2\rightarrow1$ & 50.6 & 69.8 & 50.6 & 56.6 & 54.8 & 29.4 & 54.2 & 29.4 & 0.66\tabularnewline
\hline 
Without &  &  &  &  &  &  &&&  \tabularnewline
$1\rightarrow2$ & 11.2 & 100.0 & 11.2 & 17.4 & 10.0 & 0.6 & 10.0 & 0.6 & 0.83\tabularnewline
$2\rightarrow1$ & 11.2 & 2.0 & 2.0 & 27.4 & 27.2 & 12.2 & 26.4 & 12.2 & 0.72\tabularnewline
\hline 
\end{tabular}
\par\end{centering}
\caption{Percentages of simulations with Granger-causality or CCM between x (species 1) 
and y (species 2) over 500 simulations, for the chaotic 2-species competition
model, with interactions (top rows) and
without (bottom rows).  The label both0.1 (resp. both0.2) corresponds to a combined detection criterion of p-value $<0.1$ and $\rho>0.1$ (resp. $\rho>0.2$). Similarity of causality estimates is indicated by the Sokal-Michener ($I_{SM}$) index, with both significance and effect sizes taken into account for GC and CCM (p-value $<0.1$ and $G_{x \rightarrow y}>0.04$ or $\rho>0.1$, respectively). \label{tab:PropCausality_DeterModel}}
\end{table}

CCM and GC are in general in agreement for specific simulations (i.e., specific initial conditions) corresponding to this model and parameter set: the $I_{SM}$ similarity index is close to 1, except for the weak interaction $2 \rightarrow 1$. 

\subsection*{Two-species stochastic and nonlinear dynamics}

\subsubsection*{Without environmental driver}

In our case study with two-species nonlinear competition and noise, we see that GC and CCM perform quite similarly (Table \ref{tab:Prop-causal-2speciesStoch}),
with both methods able to select properly causality in most cases ($>$95\%). CCM has slightly better rates of interactions found (no false negatives), while GC is a little more conservative, especially when considering a threshold for $G_{x \rightarrow y}$, the logarithm of the sum of squares ratio (Table \ref{tab:Prop-causal-2speciesStoch}). The false positive rate is close to the required 10\% level for both methods. Similarity indices are very close to 1, so that the two methods yield essentially similar conclusions when applied to the same time series. 

\begin{table}[H]
\begin{centering}
\begin{tabular}{c||ccc|ccccc||c}
\hline 
Method & Granger & causality &  & CCM &  &  &  & & $I_{SM}$\tabularnewline
\hline 
Thresholds & pval\textless 0.1 & $G_{x \rightarrow y}$\textgreater 0.04 & both & pval\textless 0.1 & $\rho$\textgreater 0.1 & $\rho$\textgreater 0.2 & both0.1 & both0.2 & both\tabularnewline
\hline 
With inter. &  &  &  &  &  &  & &\tabularnewline
$1\rightarrow2$ & 98.4 & 94.0 & 94.0 & 100.0 & 100.0 & 100.0 & 100.0 & 100.0 & 0.94\tabularnewline
$2\rightarrow1$ & 100.0 & 100.0 & 100.0 & 100.0 & 100.0 & 100.0 & 100.0 & 100.0 & 1.00\tabularnewline
\hline 
Without &  &  &  &  &  &  &  & \tabularnewline
$1\rightarrow2$ & 12.6 & 0.2 & 0.2 & 12.6 & 11.0 & 0.2 & 10.4 & 0.2 & 0.89\tabularnewline
$2\rightarrow1$ & 8.2 & 0.6 & 0.6 & 12.4 & 10.8 & 1.4 & 10.4 & 1.4 & 0.89\tabularnewline
\hline 
\end{tabular}
\par\end{centering}
\caption{Percentages of simulations with Granger-causality or CCM between species 1
and species 2 over 500 simulations, for the stochastic 2-species competition
model without environmental driver, with interactions (top rows) and
without (bottom rows). The label both0.1 (resp. both0.2) corresponds to a combined detection criterion of p-value $<0.1$ and $\rho>0.1$ (resp. $\rho>0.2$). Similarity of causality estimates is indicated by the Sokal-Michener index ($I_{SM}$, between 0 and 1), with both significance and effect sizes taken into account for GC and CCM (p-value $<0.1$ and $G_{x \rightarrow y}>0.04$ or $\rho>0.1$, respectively).  \label{tab:Prop-causal-2speciesStoch}}
\end{table}

The MAR($p$) model selected by BIC had a lag $p=2$ to $6$ timesteps (mean $p=4$) whenever interactions were present, $p=2$ whenever interactions were absent, confirming that small lags should be used in such models.

\subsubsection*{With an environmental driver}

The two-species model with a shared environmental driver (e.g., temperature) is considerably more complex and yields less clear cut results than stochastic two-species competition.
Regarding interactions, CCM is better at uncovering interactions that are present, as GC has a good performance for the strong interaction $2\rightarrow1$ but not the reverse interaction $1\rightarrow2$.
Both GC and CCM have difficulties indicating non-causality (when there are no interactions), and indicate false positives three to four times above the 10\% level of the test (Table \ref{tab:P-valuesStochCompetDriver}). Thresholding small effect sizes works for GC but does not solve the issue for CCM. Conditional vs pairwise GC have overall similar performance, conditional GC performs even a little worse: there is little gain in conditioning for temperature in this case. Additional examination of temperature coefficients (Appendix~\ref{sec:GC-2spdriver}) shows that poor estimation of temperature effects is to blame. 
We use seasonal surrogate time series to assess the significance of CCM, which clearly improves its power to detect interactions, but we still have spurious causalities in CCM when no interactions are present (ESM Fig. \ref{fig:Comparison-dummy-CCM}). This is therefore
a scenario where avoiding false causalities is more difficult for CCM -- though we should not forget that approximately 67-86\% of absent interactions (Table~\ref{tab:P-valuesStochCompetDriver}) are still being discovered as such by CCM, using both significance and effect size.

\begin{table}[H]
\begin{centering}
\footnotesize{
\begin{tabular}{c||ccc|ccc|ccc||c}
\hline 
Method & GC & pairwise &  & GC & conditional &  &  & CCM & &$I_{SM}$\tabularnewline
\hline 
Thresholds & pval\textless 0.1 & $G_{x \rightarrow y}$\textgreater 0.04 & both & pval\textless 0.1 & $G_{x \rightarrow y}$\textgreater 0.04 & both & pval\textless 0.1 & $\rho$\textgreater 0.2 & both & both \tabularnewline
\hline 
With &  &  &  &  &  &  &  &  &\tabularnewline
$1\rightarrow2$ & 39.4 & 13.6 & 13.6 & 45.0 & 17.8 & 17.8 & 99.6 & 98.4 & 98.4 & 0.19\tabularnewline
$2\rightarrow1$ & 97.8 & 92.2 & 92.2 & 97.8 & 90.8 & 90.8 & 100.0 & 99.8 & 99.8 & 0.91\tabularnewline
\hline 
Without &  &  &  &  &  &  &  &  & \tabularnewline
$1\rightarrow2$ & 36.2 & 10.2 & 10.2 & 42.2 & 14.4 & 14.4 & 41.8 & 36.2 & 33.0 & 0.68 \tabularnewline
$2\rightarrow1$ & 35.2 & 8.8 & 8.8 & 36.2 & 9.8 & 9.8 & 38.6 & 14.0 & 14.0 & 0.82\tabularnewline
\end{tabular}
}
\par\end{centering}
\caption{Percentages of simulations with Granger causality or CCM between species 1
and species 2 over 500 simulations for a model with 2 species and a driver
(temperature), with biotic interactions (top rows) and without (bottom rows). Similarity of causal links is indicated by the Sokal-Michener index ($I_{SM}$, between 0 and 1), with both significance and effect sizes taken into account for GC and CCM (p-value $<0.1$ and $G_{x \rightarrow y}>0.04$ or $\rho>0.2$, respectively). Causalities related to the temperature - not interactions - for CCM are shown in ESM Appendix~\ref{sec:CCM-temperature}. \label{tab:P-valuesStochCompetDriver}}
\end{table}

\subsection*{Larger interaction webs}

\begin{figure}[H]
\centering
\includegraphics[width=0.95\textwidth]{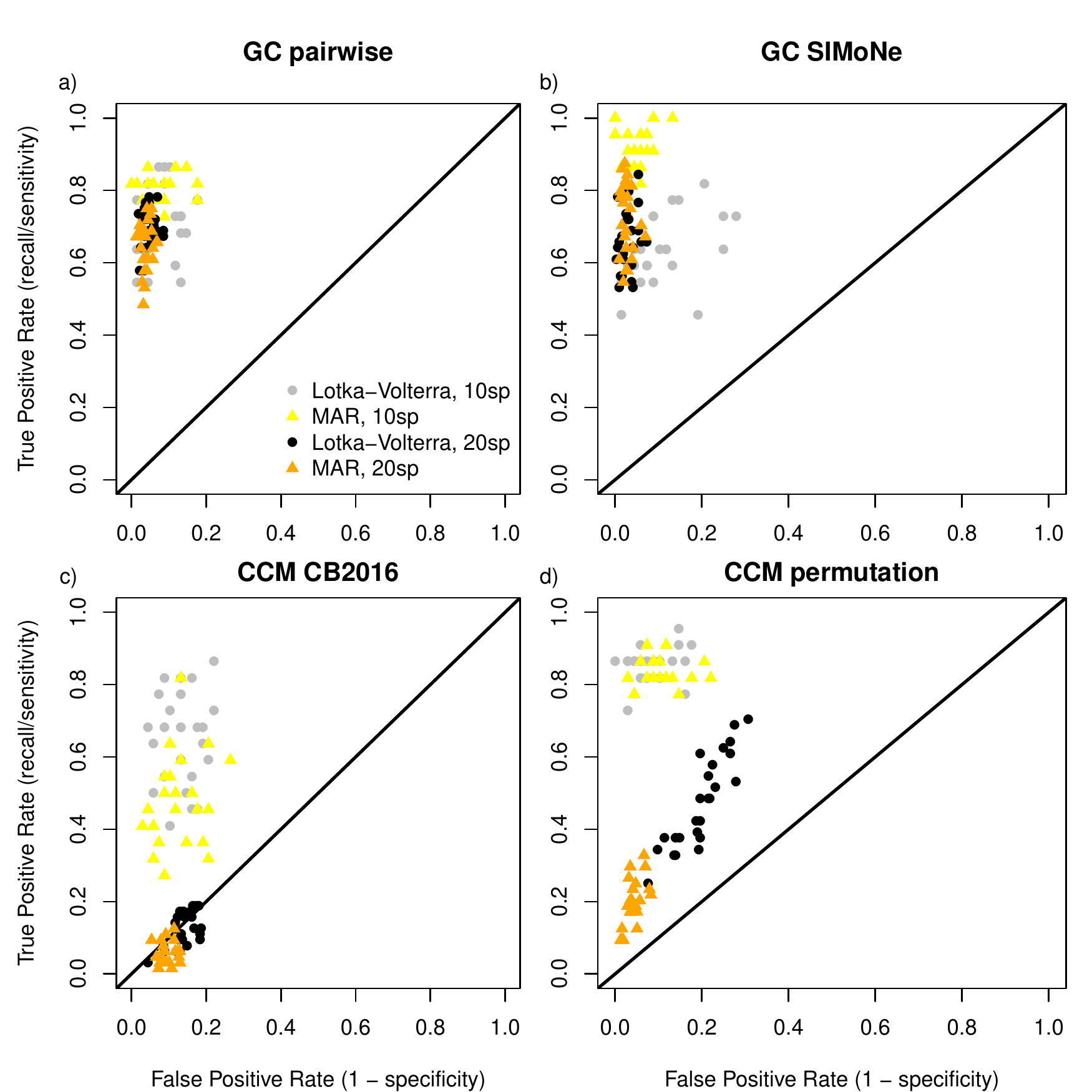}
\caption{ROC point clouds for the 10 and 20-species model using Granger Causality
(top) and CCM (bottom), with different ways of computing causality. For GC, a) corresponds to results from pairwise GC; b) is computed with structured-LASSO (SIMoNe). For CCM, we used both p-values computed as c) $\text{Pr}(\rho(L_{min})<\rho(L_{max}))$ as suggested by \cite{cobey2016limits} and d) surrogate-based p-values. In the 10-species system, one chaotic reference parameter set is considered with many initial conditions, while in the 20-species model parameters vary for each simulation. The 20-species model is a perturbed fixed point, with negative SLE. The MAR-simulated model is always the MAR(1) model obtained using the Jacobian of the Lotka-Volterra model as an interaction matrix, hence a linearization in log-scale. \label{fig:ROC-curves}}
\end{figure}

Here we report the results of analyses for 10- and 20-species modular
interaction webs. Lag order selection revealed that low-order MAR($p$)
models were selected (SI Fig. \ref{fig:Lag-order-selection_large-models-2}),
with the BIC indicating $p=1$ as the most parsimonious choice. Hence
we have focused on MAR(1) models. The high-dimensional $S\times S$
MAR(1) models include clustering (see Methods and Appendix~\ref{sec:LASSO-VAR}) because
the basic LASSO-penalized MAR(1) models poorly identify modular interaction
webs \citep{charbonnier2010weighted}.The recall, or true positive rate, that records how many actual interactions are identified as such, is on average above 70\% for the structured LASSO (\verb|SiMoNe|, \citealp{chiquet2008simone,charbonnier2010weighted})
and goes up to 80\% (90\% for the MAR model with 10 species, where it performs better than pairwise Granger causality testing). For the chaotic Lotka-Volterra model with 10-species, the method yields a little more false positives for some simulations than pairwise Granger causality (Fig.~\ref{fig:ROC-curves}a,b). For the 20-species models, where interactions are weaker and the dynamics are not chaotic, we have overall a similar performance of pairwise Granger causality testing and the structured LASSO, with a little higher average sensitivity but also a little more variance for the latter. Both methods show low false positive rates in the 20-species case, below 10\%.

\begin{figure}[H]
\centering
\includegraphics[width=0.95\textwidth]{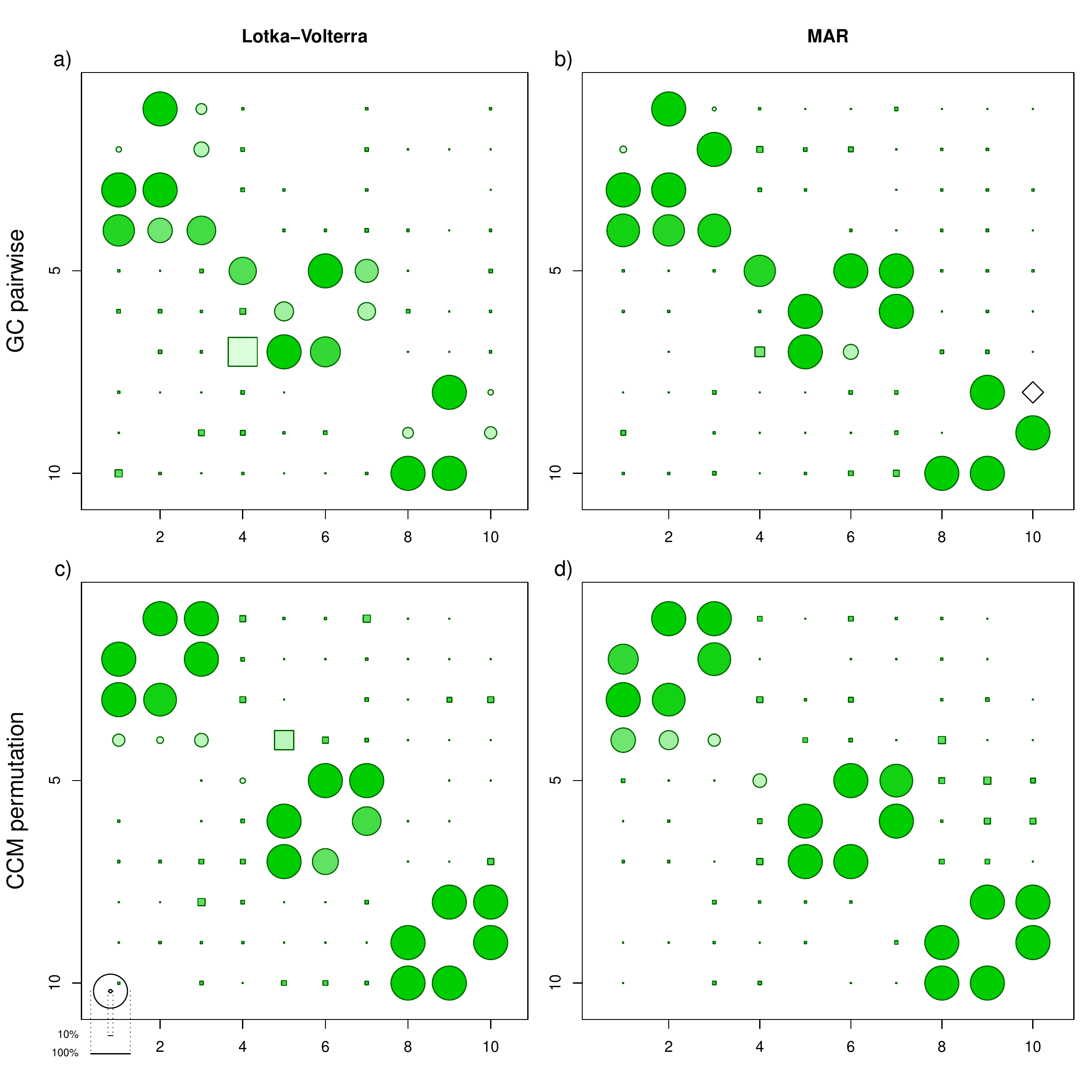}
\caption{Interaction matrices obtained from pairwise GC (top, a and b) or pairwise CCM (bottom, c and d) for 10-species communities. Green circles represent true
positives, green squares false positives, and empty diamonds false negatives. For true and false positives, the size of circles and squares
is proportional to the proportion of detection over 25 simulations. The color scale is set so that a darker green corresponds to a better performance. \label{fig:Interaction-mat-10species}}
\end{figure}

Comparing GC and CCM in ``ideal'' conditions, with the best-performing algorithms for each method (pairwise GC with a Benjamini-Hochberg correction and CCM with surrogate-based p-values) reveals that they reconstruct similar networks for the 10-species case (Fig. \ref{fig:Interaction-mat-10species}), both for the chaotic Lotka-Volterra models and the linear (MAR) approximation derived from the Jacobian matrix. Note that the MAR models have milder dynamics (no chaos), but interactions are still strong since their Jacobian matrices match those of the 10-species Lotka-Volterra, where strong interactions have been modelled. 

However, results on the 20-species model, which has weaker interactions, tend to make GC the better option, since CCM yields quite a number of false positives (Figs. \ref{fig:ROC-curves}c,d and SI Fig.~\ref{fig:Interaction-mat-20species}). This is true even with the surrogate-based p-values for CCM, that worked very well for smaller-dimensional examples. Both pairwise GC and the structured LASSO had remarkable performance in this case, and were able to make out all the network modules as well as the connecting species between them (ESM Figs. \ref{fig:Interaction-mat-20species} and Fig.~\ref{fig:Interaction-mat-20species-ter}). 

\subsection*{Summary of the results for all simulated case studies}

\begin{figure}
    \centering
    \includegraphics[width=\textwidth]{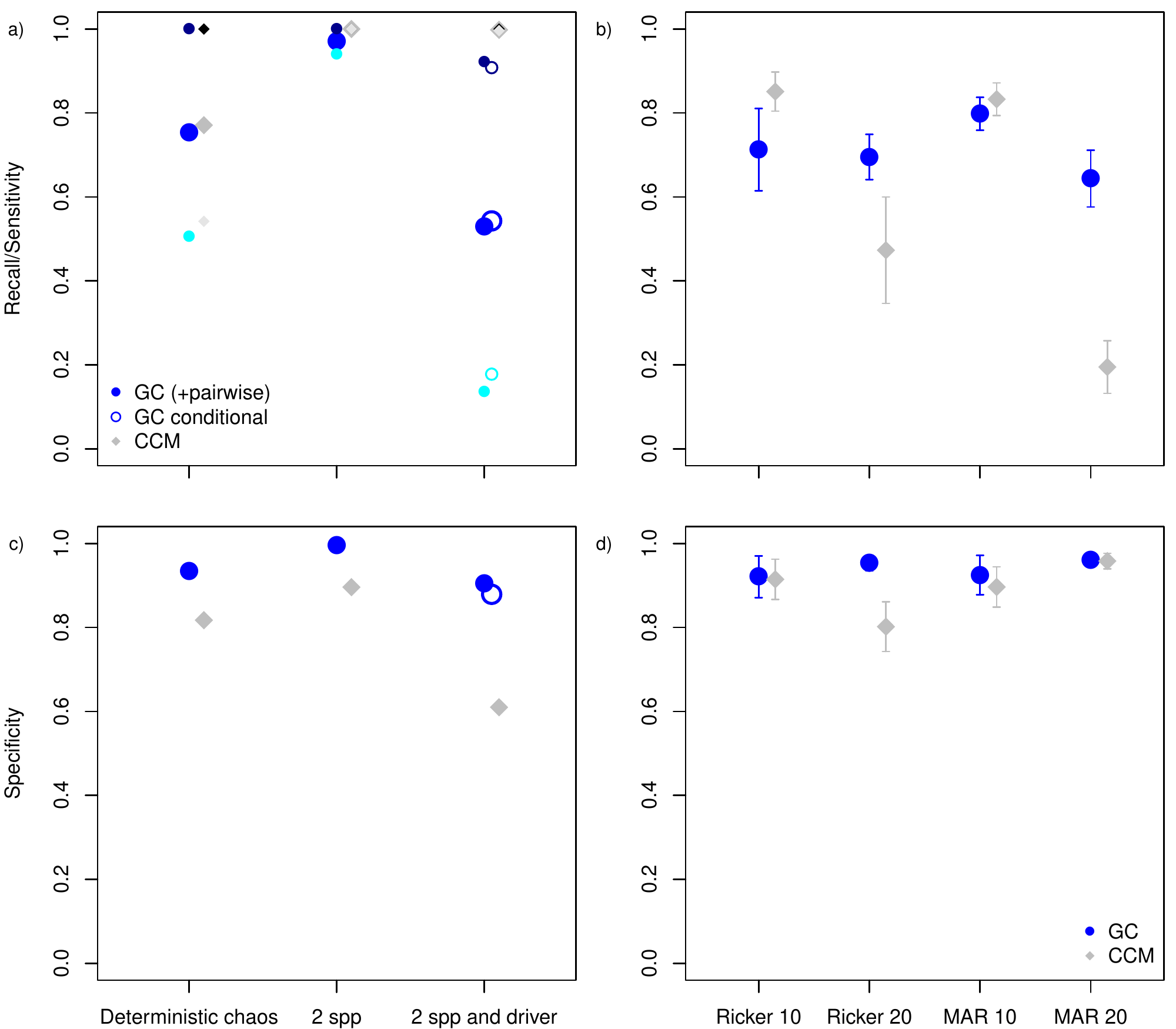}
    \caption{Recall and specificity for Granger causality (GC, blue dots) and convergent cross mapping (CCM, grey diamonds), ranked from most to least nonlinear dynamics. Causalities are considered only when they are significant at the 0.1 level (small dimension, a and c) or 0.2 level (large dimension, b and d). For small dimensionality (a and c), an additional threshold on effect sizes is considered ($G_{x\rightarrow y}>0.04$ for GC,$\rho>0.1$ for CCM). Large symbols represent the value of recall and specificity over 500 simulations for two interactions at a time for small communities. In a), the smaller symbols represent the stronger (respectively weaker) interaction in darker (resp. lighter) shade. For the 2 spp. and driver simulation, unfilled circles in a) and c) are obtained through conditional Granger causality testing. The smaller circles correspond to the weak interaction (light blue) and the strong interaction (dark blue).  For large dimensionality, error bars represent the mean (+/- SD) value of recall and specificity over 25 simulations, since the estimated network varies across simulations.
    \label{fig:diagnostics_vs_Lyapunov}}
\end{figure}

In Fig.~\ref{fig:diagnostics_vs_Lyapunov}, we present the recall or sensitivity (fraction of true positives among all positives) and the specificity (fraction of true negatives among all negatives) for all simulated case studies present in this article. The case studies are ranked by descending order of nonlinearity, from the most nonlinear model (chaotic) to the most linear, for both the low-dimensional systems considered (2 species) and high-dimensional systems (10 and 20 species). Both metrics should be close to 1 for model performance to be high; a high recall is important when the objective of a study is to find all interactions present, and a high specificity is paramount when false positives are costly. 

While we would have expected an increase in performance of GC as the dynamics are less nonlinear, combined to a decrease in performance of CCM, Fig.~\ref{fig:diagnostics_vs_Lyapunov} shows a broad overlap in the performances of both methods, both in high- and low-dimensional systems. 
CCM has slightly higher false positives rates -- lower specificity than the required $\alpha$ level -- and higher recall, except for the nonlinear 20 species systems, on which it performs worse. On the other hand, GC has difficulty finding weak interactions when there is a confounding abiotic driver. The big picture, however, is that both methods display reasonable performances in most situations. A relatively high specificity, which is a key requirement of any interaction-finding method (otherwise, the method just outputs false positives) is found in nearly all cases.

\section*{Discussion}

The purpose of this paper was to evaluate the performance of linear and parametric methods for detecting Granger causality (GC) between time series, when the underlying community dynamics are nonlinear, and to compare such performances to a nonparametric and nonlinear popular alternative, convergent cross mapping (CCM). Our main results are that linear GC, implemented using MAR($p$) models, is fairly robust to nonlinearities in ecological dynamics, when applied on the appropriate logarithmic-abundance scale and combined with model selection by information criteria. This was found to be true for all considered nonlinear simulation models, including those demonstrating chaos (Fig.~\ref{fig:diagnostics_vs_Lyapunov}). This confirms and extends findings from an investigation of the robustness to nonlinearities of log-linear MAR(1) models (with
$p$ restricted to 1 lag, \citealp{certain2018how}).

Comparison to the CCM framework by \citet{sugihara2012detecting} further revealed that CCM and MAR($p$) / Granger causal modelling can in fact - surprisingly - yield relatively similar results in nonlinear and stochastic dynamical systems of interacting species. Evidence for this comes \emph{both}
from highly nonlinear systems for which CCM and GC infer the same interactions (deterministic chaos, stochastic competition) and from cases where both methods seem to
fail to some degree, such as two competing species forced similarly by a shared environmental driver, where false positives are frequent (between 15\% and 40\%) for CCM and false negatives frequent for GC (one causal direction missed). Therefore, an important conclusion from our study is that both Granger causality and CCM yield mostly similar inferences on similar datasets, with only a couple of exceptions.  

Moreover, we used here false discovery rate corrections and regularized models (i.e., LASSO-penalized models for modular interaction networks, \citealp{charbonnier2010weighted}) to tackle relatively-high dimensional systems (10 and 20 species). This allows to better infer Granger causality in lifelike contexts that, we surmise, will be most exciting to ecologists working on interacting species using community-level data. Our results demonstrated that simple pairwise Granger causality (i.e., using $2\times2$ MAR($p$) models many times with a correction for multiple testing) can be as good as the penalized MAR(1) models in finding the interaction network. We elaborate on these results and possible explanations below.

\subsection*{Can Granger causality be applied to highly nonlinear coupled dynamical systems?}

Our results showed that Granger causality, in its log-linear form, is robust to the presence of nonlinearities in the underlying multivariate dynamical system. Nonlinear variants of Granger causality \citep{Marinazzo2008,yang2017reconstruction,hannisdal2018causality} can also be used to infer interactions in nonlinear and stochastic dynamical systems. These conclusions are further supported by the neuroscience literature \citep{Ding2006,Chen2006,Marinazzo2008,papana2013simulation,barnett_mvgc_2014} which, unlike ecology, commonly uses Granger causality on nonlinear (and stochastic) dynamical systems.  

Ecologist views over Granger causality have likely been shaped by the influential paper of \citet{sugihara2012detecting}, who suggested that Granger causality would
work well for simulated (log)-linear systems (which they referred to as ``stochastic'') while CCM would work
well for near-deterministic nonlinear dynamical systems. Given the history of both techniques, this makes intuitive sense. However, our tests on simulated data revealed that the domains of applicability of both techniques overlap to a great extent. Several differences between our analyses and those performed by \citet{sugihara2012detecting}
(in their Supplementary Material) allow to explain this greater overlap, which we develop here.

First, Granger-causality analyses performed by \citet{sugihara2012detecting} on the Veilleux and other datasets rely on a slightly dated model selection procedure (pre-information criteria) which produced overparameterized autoregressive models with very long lags (e.g., $p>10$). Here, re-analysing the data with a more classic, information-criteria motivated lag order
selection, we have shown that GC is perfectly able to find causality in the classic Veilleux \textit{Paramecium-Didinium} predator-prey datasets, that were used as a key nonlinear example by \citet{sugihara2012detecting}.

Second, we sampled many chaotic simulated datasets, corresponding to many initial conditions. Although some chaotic datasets may be difficult to identify for GC techniques, these are very few, as MAR($p$) models and GC inference detected above 95\% of the strong true interactions (where CCM found 100\%), while both methods detected 50\% of weak true interactions, in two-species chaotic models. This result was quite unexpected, as we thought that CCM would completely dominate the scores. Thus GC testing can be useful for highly nonlinear systems, even chaotic ones, and it additionally tends to produce correct rates of false positives when the null hypothesis of no interactions is true (an important aspect as well).

Third, we found that data simulated with log-linear autoregressive models can also be well-identified by CCM, in both low and high dimensions, even though CCM relies upon the possibility to reconstruct an attractor in state space. This is further proof of the overlap between the domains of applicability of linear GC and CCM. Incidentally, it is also proof that the assumption of a deterministic dynamical system, often seen as a pre-requisite for CCM \citep{runge2019inferring,langendorf2019can}, may in fact not be needed.

\subsection*{How can Granger causality and convergent cross mapping yield similar inferences, in spite of seemingly opposite assumptions?}

Here, we would like to go back to the heart of the issue that \citet{sugihara2012detecting} highlighted, i.e., ``causality reversion'' in nonlinear dynamical systems. Note that while we offer some suggestions as to how GC and CCM performances can overlap to a large extent, we have no definite answer as to why (mathematically speaking); more research is needed on that point. 

The standard Granger causality concept holds that whenever a probabilistic model $Y_{t+h}|({Y_{k},X_{k}})_{k\in \llbracket t-p+1,t \rrbracket}$
better predicts the observed time series $\mathbf{y}=(y_t)_{t \in \llbracket 1,T \rrbracket}$ than a model $Y_{t+h}|{(Y_{k})}_{k\in \llbracket t-p+1,t \rrbracket}$,
then ``$x$ is causal for $y$''. Most often the time horizon for prediction considered is $h=1$, which is the perspective adopted in this article. CCM instead holds that causality flows from $x$ to $y$ whenever $\rho(\mathbf{x},\hat{\mathbf{x}}|M_{Y})$ increases strongly
with the library size $L$ used to reconstruct $\mathbf{x}$ from the shadow manifold $M_{Y}$. It seems that in the latter method, $x$ causes $y$ whenever knowledge about $y$ can be used to reconstruct $x$. However, verbal reasoning is treacherous there. To determine whether
$x$ causes $y$:
\begin{itemize}
\item GC compares knowledge about $Y_{t}$ vs. knowledge about $X_{t},Y_{t}$ in prediction of $Y_{t+1}$
\item CCM compares knowledge about $M_Y$ vs. no knowledge about $M_Y$ in prediction of $X_{t}$.
\end{itemize}
There is no direct conditionality upon past $X_{t}$ values
in the prediction step of the algorithm for CCM. Thus, there seems to be no causality reversion that is intrinsic to nonlinear dynamic testing: GC and CCM are simply two different types of causal inference that are based upon \emph{different assumptions on the conditioning set} and \emph{ways to select models}. 

Finally, GC and CCM both try, as do other approaches based on continuous-time stochastic processes and martingale theory \citep{commenges2009general,aalen2012causality}, to reconstruct a stochastic dynamical system where interactions are defined as influences of state variables upon the rates of change of the system. No matter how different their historical origins are, GC and CCM are bound to exhibit some similarities because they define interactions in a similar manner. 

\subsection*{Inferring interactions under environmental forcing}

We observed a lowered performance of linear GC in detecting interactions (whenever these were present) in the case of competition with a shared abiotic driver (temperature, which was both autocorrelated and seasonal). Notably, the weak interaction was entirely missed. This seems to be partly due to poor temperature effect estimates. Indeed, there was a bias in the temperature effect estimates (underestimation, Appendix~\ref{sec:GC-2spdriver}), that was still present when taking only statistically significant estimates. Additional analyses (in our \verb|2species_driver| code folder, \citealt{barraquand_picoche_2020}) have demonstrated that better temperature effect estimates could be obtained if the temperature consisted of white noise or if the underlying deterministic skeleton of the model was more linear (stable fixed point obtained by lowering intrinsic growth rates). This mirrors results of \citet{certain2018how} who found a good estimation of the temperature effect on growth rates (on one species only) for a stable fixed point in competition models. The fact that we have a good estimation of causal effects with nonlinear models forced by a white noise driver is consistent with our study of the two-species Ricker competition model without the environmental driver: adding a white noise driver amounts to increasing the level of background stochasticity, not its nature. Hence it seems to be the interaction between a strongly autocorrelated forcing noise and a nonlinear (albeit nonchaotic) stochastic dynamical system that made the estimation challenging for linear GC here. This was true with and without interactions between the species, leading to false negatives, as stated above, but also false positives (although these remained around 10\% when using cutoffs on effect size). Because the temperature was seasonal, using a seasonal dummy variable helped to improve temperature estimates, but they still were quite variable between simulations. This should not be overly surprising: combined estimation of nonlinear or delayed density-dependence and environmental effects under autocorrelated environmental forcing is a known challenge for parametric methods \citep{jonzen2002irreducible,jiang2003autocorrelated,linden_mischaracterising_2013}. 
Non-parametric CCM was found in this particular scenario to avoid well false negatives. However, it did introduce a sizeable number of false positives (between 40\% and 15\%) which suggests that the method could be better calibrated, a topic that we tackle below. 

\subsection*{Issues in calculating p-values and confidence intervals}

So far, we mainly discussed the performance of CCM and GC in terms of sensitivity and specificity. Such unit-less interpretation of the model outputs dominates benchmarks and tests in the ecology and physics literature (e.g., see recently
\citealp{krakovska2018comparison}). This implicitly requires setting cut-offs for p-values and effect sizes to decide when an interaction is present. This is possible in a simulation context, but in practice, p-values, Bayes factors, and confidence intervals are the quantities that are typically reported. Therefore, a question of interest is: do GC and CCM consistently produce precise p-values and confidence intervals? 
Our results show that while overall GC and CCM produce sound results, statistical indicators for both methods are not always very well calibrated. This is exemplified by all the Tables in our manuscript in that, in the case where the null hypothesis of no interactions is true, the percentage of p-values below 10\%, for both GC and CCM, does not always match exactly the 10\% level of the test employed (although GC fares usually a bit better than CCM in this regard).

The absence of proper calibration for GC is easily explained by the fact that the model that generated the data (nonlinear) and the model used to analyze it (log-linear) are not the same, and thus there is no reason to expect perfectly calibrated p-values. Our results support other findings that confidence intervals for MAR(1) models, when fitted to the data
generated by more nonlinear models, tend to be `too narrow' \citep{certain2018how},
in the sense that there is poor coverage of the point estimate. Nonlinear
Granger causality methods (\citealp{schreiber2000measuring},
\citealp[see][for reviews)]{Palus2008,amblard2013relation,papana2013simulation}, could be of use to improve causality detection by obtaining more exact p-values. Transfer entropy (SI Appendix~\ref{sec:Nonlinear-Granger-causality}), in particular, admits linear GC as a special case \citep{barnett2012transfer}, and therefore provides an interesting bridge to classical MAR($p$) modelling. 

Why p-values for CCM were also imperfectly calibrated is unclear. While the original CCM article \citep{sugihara2012detecting} method did not directly calculate p-values, further work has recommended to use surrogate time series to do so \citep[e.g.,][]{deyle2016global,ye2018redm}, a suggestion with which we concur. \citet{cobey2016limits}
proposed another method to calculate p-values for CCM based on the increase of $\rho$ with library size $L$: although this technique made sense in theory, it was not found to work well in practice (see SI section~\ref{subsec:Choice-of-p-values}). We therefore considered several surrogate-based p-values and chose the best-performing ones (Figs. \ref{fig:Comparison-pval-CCM}, \ref{fig:Comparison-of-false-neg-pvalues}
in Appendices), but in some cases -- with a confounding abiotic driver or with 20 species -- this was not completely satisfactory. Formal statistical inference for CCM could therefore be improved. Another idea could be to combine both worlds and perform surrogate-based nonlinear Granger-causality inference \citep{schreiber2000measuring,schreiber2000surrogate,Palus2008}.

\subsection*{The specific challenges of high-dimensional, many species interacting systems}

We found here that both GC and CCM were scalable to large interaction networks (10 or 20 species) for relatively long time series by ecological standards (i.e., 300 to 700 time steps). We used both false discovery rate corrections and regularized models (i.e., LASSO-penalized MAR(1) models developed for modular interaction networks, \citealp{charbonnier2010weighted}). 
 
One surprising find, for Granger analyses, was that the structured LASSO did not massively outperform the FDR-corrected pairwise analyses (since the underlying true network is specified from conditional not pairwise interactions). One way to interpret these results is in terms of correcting for confounders vs collider bias. Fitting a high-dimensional model, even with regularization through the LASSO, has the benefits of including in the estimation of an interaction $j \rightarrow i$ the other (potentially many) interacting species. This can be construed as correcting for potential confounders when estimating an interaction. However, any incorrectly included species in the network can generate what is known in causal inference as collider bias \citep{pearl2009causal}: if species 6 does not truly affect species 1, but is included in the dynamical model for species 1, then the effect $2 \rightarrow 1$ might be poorly estimated. Therefore, there is a trade-off between accounting for confounding factors and avoiding collider bias. Pairwise FDR-corrected analyses seemed to realize the best-trade off for the chaotic 10 species Lotka-Volterra models, though not for the 10 species MAR models for which the LASSO was better. This could suggest that when the model functional forms are exact (in the MAR case), avoiding collider bias is easy, but when these are poorly known and the dynamics are highly nonlinear, avoiding collider bias is harder. 
 
 However, the performances of the structured LASSO MAR(1) models may be diminished by other choices: (1) we used MAR(1) not MAR($p$), which limits the ability of the autoregressive model to mimic the nonlinear system and (2) we did not use iterative model fitting, e.g., using the first inferred network as prior for the latent structure or as initial condition for further estimates. Both of these ideas may improve the network inference. One reason why we used MAR(1) modelling for the high-dimensional systems, outside of just simplicity, was that $p=1$ was selected in the 20-species case based on BIC, SI section~\ref{sec:lagorder-HDsystems}. But this selection of the lag itself did not use regularization. Selecting both the interaction matrix sparsity and the lag order through regularization in high-dimensional MAR($p$) modelling is extremely challenging, because there are many ways to connect the number of time lags $p$ to the LASSO penalties \citep{michailidis2013autoregressive,nicholson2017bigvar}. \citet{mainali2019detecting} recently used the R package \verb|BigVAR| \citep{nicholson2017bigvar}, a promising technique for penalized MAR($p$) fitting, but few interactions were found and model performance was not evaluated with simulations; we have found here in contrast that without a latent network structure, unstructured LASSO-based methods perform poorly on large networks. Hence our choice of \verb|SIMoNe| \citep{charbonnier2010weighted,chiquet2008simone} which sticks to $p=1$ but allows the latent structure to be specified as a stochastic block model (\citealp{daudin2008mixture}, see ESM Appendix~\ref{sec:LASSO-VAR} for details). Combining structured LASSO modelling with models more sophisticated than MAR(1) remains an area where development is needed. 


\subsection*{Causal inference for nonlinear and stochastic ecological systems: going further}

Overall, both linear Granger causality and convergent cross mapping can show good recall (sensitivity) and specificity for highly nonlinear and stochastic dynamical systems. Their domains of applicability overlap to a great degree. Rather than choosing one of these frameworks based on the supposed degree of nonlinearity or stochasticity of the ecological system studied \citep[e.g.,][]{runge2019inferring}, we suggest that which one to use may be decided based on the goal and constraints of the analysis. 

For instance, (log)-linear GC, being a fully parametric framework, can easily be extended to situations where we have small counts that preclude data transformation. This requires using a log-link function rather than an actual log transformation, as in Poisson Log-Normal models \citep[e.g.,][]{chiquet2018variational} and other flavours of latent variable modelling \citep[e.g.,][]{warton2015so,ovaskainen2017make}. It may likewise be very useful when one wants to introduce compositionality constraints in microbiome studies \citep{bjork2017dynamic}. Conversely, convergent cross mapping or nonlinear Granger causality techniques allow for a much finer reconstruction of the attractor shape, which can be very useful to compare to the attractor shape of candidate mechanistic models (e.g., coupled differential equations models), if those exist. In some cases, using both frameworks, to increase the robustness of the interaction inference, is another idea \citep{hannisdal2017common}. 

Looking at the various implementations of GC and CCM, it seems that the most critical methodological choices are rarely located along a ``linear vs nonlinear model'' gradient, but instead boil down to two characteristics. First, details matter: faulty selection of the lag order $p$ of autoregressive models results in nonsensical GC inference, and yet proper $p$ selection yields causal inferences fairly robust to nonlinearities. Likewise, versions of CCM including significance testing are quite sensitive to the p-value definition, and surrogate-based p-values should be preferred. In other words, the devil is always in the details of the test or model selection, for Granger-based or CCM-based methods alike. Second, a key choice to make is what constitutes the ``conditioning set'', i.e., the variables that are known to be important confounders and are \textit{de facto} included in the time series model \citep{eichler2013causal}. For instance, an unknown confounder such as seasonal temperature or an invading species can massively thwart any attempt at interaction inference if not corrected for. And even when corrected for (i.e., adding the confounder to the autoregression or considering surrogate time series), this is the scenario where we observed the largest proportion of false positives for both GC and CCM. Strategies to better understand how to choose and handle this conditioning set when performing causal inference will be, we believe, a very important feature of ecological interaction inference for the years to come. Several algorithms have been already put forward \citep{eichler2013causal,runge2018causal,runge2019inferring,runge2019detecting}, and much remains to be done to better incrementally select variables in order to assemble networks. 

\section*{Acknowledgements}

FB thanks Julien Chiquet and Camille Charbonnier for advice on the structured LASSO and Grégoire Certain for discussions on MAR modelling. FB and CP were supported by the French ANR through LabEx COTE (ANR-10-LABX-45). We thank Ethan Deyle, Adam Clark and Hao Ye for feedback, notably regarding surrogate time series testing for CCM. Constructive referee suggestions improved the manuscript, especially the figures. 

\section*{Author contributions}

All authors contributed to the project design. FB and CP constructed the case studies, wrote the computer code, and analysed the real and simulated data. All authors contributed to the interpretation of the results. FB wrote a first draft of the manuscript, which was then edited by all authors. 

\section*{Data accessibility}

Codes for the analyses presented in this paper are available at \url{https://github.com/fbarraquand/GCausality} and are published at Zenodo with DOI \href{http://doi.org/10.5281/zenodo.3967591}{10.5281/zenodo.3967591} \citep{barraquand_picoche_2020}. 

\bibliographystyle{ecol_let}
\bibliography{GC_bib,GC_2}

\clearpage

\part*{Electronic Supplementary Material}

\textbf{Appendices} for \textit{Inferring species interactions using Granger causality and convergent cross mapping} by Barraquand F., Picoche C., Detto M. and Hartig F. DOI: 10.1007/s12080-020-00482-7

\renewcommand\thesection{S\arabic{section}}
\renewcommand\thefigure{S\arabic{figure}}
\setcounter{figure}{0}   

\renewcommand\theequation{S\arabic{equation}}
\setcounter{equation}{0}   

\section{Extensions of Granger causality}

\subsection{LASSO-based MAR(1) models}\label{sec:LASSO-VAR}

\paragraph{Classic MAR(1) estimation} 

We follow here the presentation of \citet{charbonnier2010weighted,chiquet2008simone}, with some notational adaptations from \citet{ives2003ecs} and keep our notations in line to those of the main text. We start with the MAR(1) model without external input for the log-abundance vector $\mathbf{x}_t$, which we assume to be scaled and centered. The model is given by 

\begin{equation}
\mathbf{x}_{t}=\mathbf{Bx}_{t-1}+\mathbf{e}_{t},\mathbf{e}_{t}\sim\mathcal{N}_{d}(\mathbf{0},\mathbf{D} )\label{eq:MAR1_charbonnier}
\end{equation}

where matrix $\mathbf{B}$ has dimension $d \times d$, same as in the main text, and $\mathbf{D}_{ii} = \sigma_i^2$ with $\mathbf{D}$ a diagonal matrix. The model is observed for times $t=1,...,T+1$ which then defines a $T \times d$ matrix of observed densities $\mathbf{X} = [ \mathbf{x}_1,\mathbf{x}_1,...,\mathbf{x}_{T} ] '$ (the prime denotes matrix transposition) and a $T \times d$ matrix of densities observed just one time step after  $\mathbf{Y} = [ \mathbf{x}_2,\mathbf{x}_3,...,\mathbf{x}_{T+1} ] '$. 

\citet{charbonnier2010weighted} actually study a slightly different representation of the model 
\begin{equation}
\mathbf{x}_{t}'=\mathbf{x}_{t-1}' \mathbf{A} +\mathbf{e}_{t}',\mathbf{e}_{t} \sim\mathcal{N}_{d}(\mathbf{0},\mathbf{D} )\label{eq:MAR1_charbonnier_original}
\end{equation}
where $\mathbf{A} = \mathbf{B}'$ is the transpose of $\mathbf{B}$ and $\mathbf{x}_{t}'$ is now a line vector. This representation may be familiar to readers acquainted with discrete-time Markov chains, although it is less frequent for VAR(1) models. 

As remarked by \citet{charbonnier2010weighted}, this model can be fitted to data by using the following relations:
\begin{itemize}
    \item $\mathbf{S} = \frac{1}{T} \mathbf{X}' \mathbf{X}$ is the empirical variance-covariance matrix
    \item $\mathbf{V} = \frac{1}{T} \mathbf{X}' \mathbf{Y}$ is the temporal autocovariance matrix
\end{itemize}

Optimizing the log-likelihood of the MAR(1) process is then equivalent to 
\begin{equation}\label{eq:LL-MAR1}
    \underset{\mathbf{A}}{\text{max}} \{ \text{Tr} (\mathbf{V}'\mathbf{A}) - \frac{1}{2} \text{Tr} (\mathbf{A}' \mathbf{S} \mathbf{A}) \}
\end{equation}

The solution of this maximization problem is then given by $\mathbf{A}^{\text{mle}} = \mathbf{S}^{-1} \mathbf{V}$. 
This is proved by (1) reducing eq.~\ref{eq:LL-MAR1} to an OLS problem and (2) compute $\mathbf{A}^{\text{ols}}$ as $(\mathbf{X}' \mathbf{X})^{-1} \mathbf{X}' \mathbf{Y} = \mathbf{S}^{-1} \mathbf{V}$. This solution requires that $\mathbf{S}$ is invertible, which requires $T<d$. 
For $T<d^2$ some degree of regularization will be needed as well. We get back to our original formulation doing $\mathbf{B} = \mathbf{A}'$. 

\paragraph{LASSO-based $A$ estimate} 
Sparsity can be enforced with a classical $L_1$ penalty, so that 

\begin{equation}\label{eq:LL-MAR1-L1}
    \underset{\mathbf{A}}{\text{max}} \{ \text{Tr} (\mathbf{V}'\mathbf{A}) - \frac{1}{2} \text{Tr} (\mathbf{A}' \mathbf{S} \mathbf{A}) - \rho || \mathbf{A}||_1\}
\end{equation}

Unfortunately, this tends to (a priori) penalize all coefficients alike, and therefore to consider by default that the network has no structure. We have always found this method to lead to worse results than those assuming some structure, for modular ground truth networks like those considered in our 10- and 20-species simulations. 
When the network is structured, one can introduce a latent structure by assuming that a network $\mathcal{P}$ is structured into $\mathcal{Q}$ classes. We note $Z_{iq}$ the indicator function (a random variable) whose value is $1$ if species $i$ belongs in class $q$ (this can be a network module, for instance). 
The choice of the latent structure follows \citet{ambroise2009inferring}, who use the mixture framework of \citet{daudin2008mixture}. A Laplace distribution on the network weights is chosen, a Laplace prior on coefficients being equivalent to LASSO optimization \citep{tibshirani2015statistical}. It is therefore assumed that the a priori link strength between species $i$ and species $j$ is distributed as 
\begin{equation}
f_{ijql}(x) = \frac{1}{2 \lambda_{ql}} \exp \left(- \frac{|x|}{\lambda_{ql}} \right)
\end{equation} 
where $\lambda_{ql}$ describe the intensity of the link between classes $q$ and $l$. 

Implementing this prior on the interaction strength then equates to the following optimization problem for the likelihood $\mathcal{L}$ with latent network structure $\mathbf{Z}$. 

\begin{equation}\label{eq:LL-MAR1-L1-latent}
    \hat{\mathbf{A}} = \text{arg} \text{max} \log \mathcal{L} (\mathbf{Y}, \mathbf{A}; \mathbf{Z}) =  \text{arg} \text{max} \left\{ \text{Tr} (\mathbf{V}'\mathbf{A}) - \frac{1}{2} \text{Tr} (\mathbf{A}' \mathbf{S} \mathbf{A}) - || \mathbf{P}^{\mathbf{Z}} \star \mathbf{A}||_1\right\}
\end{equation}

where $\mathbf{P}^{\mathbf{Z}} = \left( {P_{ij}}_{i,j \in \mathcal{P}}^\mathbf{Z} \right) = \sum_{q,l \in \mathcal{Q}} \frac{Z_{iq}Z_{jl} }{\lambda_{ql}}$ are the penalties encapsulating the network structure (see \citealp{ambroise2009inferring} for details on such penalties). 
We refer to \citet{charbonnier2010weighted} for the details of the algorithm used here. In essence the particular structure of the model allows to reduce this global LASSO optimization to $d$ LASSO-style problems, which makes it much faster. The tuning of the penalty parameter is then done using BIC \citep{charbonnier2010weighted}. 

\subsection{Example simulation of a 2-species stochastic Ricker model}\label{sec:2sppRickerillustration}

\begin{figure}[H]
\begin{centering}
\includegraphics[width=0.8\textwidth]{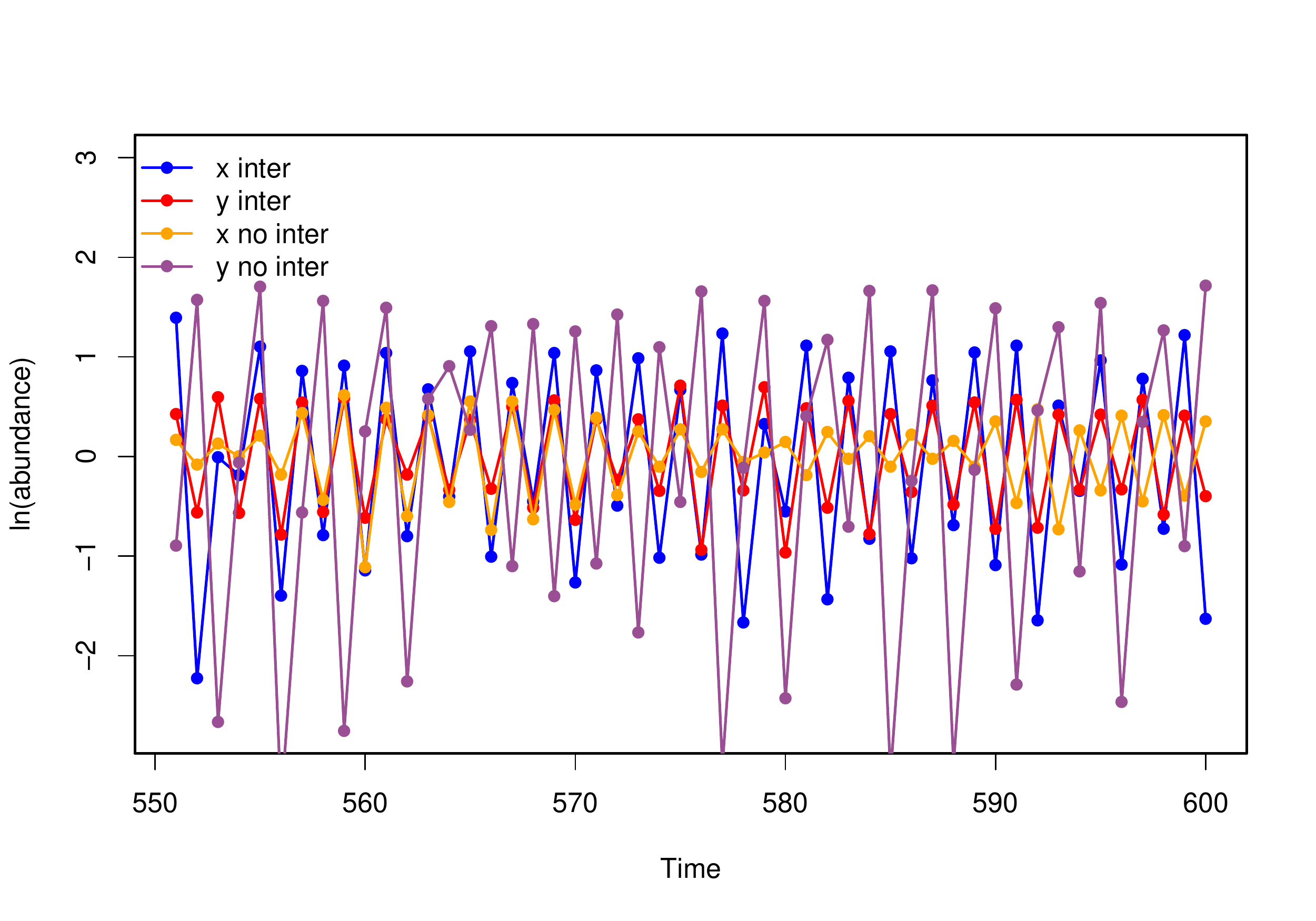}
\par\end{centering}
\caption{Abundance time series for a 2-species Ricker model (eqs.~\ref{eq:twoSpeRickerStoch_compet}--\ref{eq:twoSpeRickerStoch_compet_2}) with (blue =
species 1, red = species 2) and without (purple = species 1, orange = species 2) competition.}
\end{figure}

\subsection{From Lotka-Volterra to multivariate autoregressive model}\label{sec:jacobian}

Our objective here is to keep a commensurate interaction matrix between the Lotka-Volterra model and its corresponding log-linearised autoregressive version. We therefore need to compute the corresponding Jacobian matrix $\mathbf{J}$ (see also \citealp{ives2003ecs,certain2018how}). The Lotka-Volterra model can be written, after log-transformation and centering, as:

\begin{eqnarray}
    \mathbf{n}_{t+1}&=&\mathbf{n}_t+\mathbf{A} \mathbf{N}_t + w_t, w_t\sim \mathcal{N} (\mathbf{0},\mathbf{\Sigma})\\
    \Rightarrow n_{i,t+1}&=&f_i(n_{k,t})_{k\in[1,S]}+\epsilon_{i,t}, \epsilon_{i,t}\sim  \mathcal{N} (0,\sigma')
\end{eqnarray}
where $\mathbf{N}_t=(e^{n_{1,t}},..,e^{n_{S,t}})^T$ and $f_i(\mathbf{n})=n_i+a_{i\bullet}\mathbf{N}_t$, with $a_{i\bullet}\mathbf{N}_t=\sum_{k=1}^Sa_{ik}N_{k,t}$.\\

We can write the Jacobian matrix elements as $J_{ij}= \frac{\partial f_i}{\partial n_j}$. Then,

\begin{eqnarray}
J_{ij}&=&\frac{\partial n_i}{\partial n_j}+\sum_{k=1}^Sa_{ik}\frac{\partial e^{n_k}}{\partial n_j}\\
J_{ij}&=&\delta_{ij}+a_{ij}e^{n_j} = \delta_{ij}+a_{ij}N_j
\end{eqnarray}

\subsection{20-species model interaction matrix}\label{sec:20sp-matrix}

The 20-species model has an interaction structure  
that is still fairly modular (eq.~\ref{eq:mat_20species}) yet
some species act as links between the different modules (e.g., species
4 and 5).

\begin{equation}
\chi =\left(\begin{array}{cccccccccccccccccccc}
1 & 1 & 1 & 0 & 0 & 0 & 0 & 0 & 0 & 0 & 0 & 0 & 0 & 0 & 0 & 0 & 0 & 0 & 0 & 0\\
1 & 1 & 1 & 0 & 0 & 0 & 0 & 0 & 0 & 0 & 0 & 0 & 0 & 0 & 0 & 0 & 0 & 0 & 0 & 0\\
1 & 1 & 1 & 0 & 0 & 0 & 0 & 0 & 0 & 0 & 0 & 0 & 0 & 0 & 0 & 0 & 0 & 0 & 0 & 0\\
1 & 1 & 1 & 1 & 1 & 0 & 0 & 0 & 0 & 0 & 0 & 0 & 0 & 0 & 0 & 0 & 0 & 0 & 0 & 0\\
0 & 0 & 0 & 1 & 1 & 1 & 1 & 0 & 0 & 0 & 0 & 0 & 0 & 0 & 0 & 0 & 0 & 0 & 0 & 0\\
0 & 0 & 0 & 0 & 1 & 1 & 1 & 0 & 0 & 0 & 0 & 0 & 0 & 0 & 0 & 0 & 0 & 0 & 0 & 0\\
0 & 0 & 0 & 0 & 1 & 1 & 1 & 0 & 0 & 0 & 0 & 0 & 0 & 0 & 0 & 0 & 0 & 0 & 0 & 0\\
0 & 0 & 0 & 0 & 0 & 0 & 0 & 1 & 1 & 1 & 1 & 1 & 1 & 0 & 0 & 0 & 0 & 0 & 0 & 0\\
0 & 0 & 0 & 0 & 0 & 0 & 0 & 1 & 1 & 1 & 1 & 1 & 1 & 0 & 0 & 0 & 0 & 0 & 0 & 0\\
0 & 0 & 0 & 0 & 0 & 0 & 0 & 1 & 1 & 1 & 1 & 1 & 1 & 0 & 0 & 0 & 0 & 0 & 0 & 0\\
0 & 0 & 0 & 0 & 0 & 0 & 0 & 1 & 1 & 1 & 1 & 1 & 1 & 0 & 0 & 0 & 0 & 0 & 0 & 0\\
0 & 0 & 0 & 0 & 0 & 0 & 0 & 1 & 1 & 1 & 1 & 1 & 1 & 0 & 0 & 0 & 0 & 0 & 0 & 0\\
0 & 0 & 0 & 0 & 0 & 0 & 0 & 1 & 1 & 1 & 1 & 1 & 1 & 0 & 0 & 0 & 0 & 0 & 0 & 0\\
0 & 0 & 0 & 0 & 0 & 0 & 0 & 0 & 0 & 0 & 1 & 1 & 1 & 1 & 1 & 0 & 0 & 0 & 0 & 0\\
0 & 0 & 0 & 0 & 0 & 0 & 0 & 0 & 0 & 0 & 0 & 0 & 0 & 1 & 1 & 1 & 1 & 0 & 0 & 0\\
0 & 0 & 0 & 0 & 0 & 0 & 0 & 0 & 0 & 0 & 0 & 0 & 0 & 0 & 1 & 1 & 1 & 0 & 0 & 0\\
0 & 0 & 0 & 0 & 0 & 0 & 0 & 0 & 0 & 0 & 0 & 0 & 0 & 0 & 1 & 1 & 1 & 0 & 0 & 0\\
0 & 0 & 0 & 0 & 0 & 0 & 0 & 0 & 0 & 0 & 0 & 0 & 0 & 0 & 0 & 0 & 0 & 1 & 1 & 1\\
0 & 0 & 0 & 0 & 0 & 0 & 0 & 0 & 0 & 0 & 0 & 0 & 0 & 0 & 0 & 0 & 0 & 1 & 1 & 1\\
0 & 0 & 0 & 0 & 0 & 0 & 0 & 0 & 0 & 0 & 0 & 0 & 0 & 0 & 0 & 0 & 0 & 1 & 1 & 1
\end{array}\right)\label{eq:mat_20species}
\end{equation}

\subsection{Transfer entropy and nonlinear Granger causality}\label{sec:Nonlinear-Granger-causality}
Transfer entropy can be defined as

\begin{equation}
{\mathcal{T}}_{x\rightarrow y|z}=H(\mathbf{y}^{T+1}|\mathbf{y}^{T},\mathbf{z}^{T})-H(\mathbf{y}^{T+1}|\mathbf{y}^{T},\mathbf{x}^{T},\mathbf{z}^{T})
\end{equation}

where $\mathbf{y}^{T+1}=(y_{2},...,y_{T+1})$ and $\mathbf{y}^{T}=(y_{1},...,y_{T})$ and $\mathbf{x}^T,\mathbf{z}^T$ are similarly defined.The quantity $H(x|y)=H(x,y)-H(y)$ is a conditional entropy, defined
with $H(x)$ the Shannon entropy. It has then been shown that the
Granger causal measure $G_{x\rightarrow y|z}=\ln(\frac{\sigma_{\eta}^{2}}{\sigma_{\epsilon}^{2}})$
where the residuals errors are taken from eqs. \ref{eq:directGC} --\ref{eq:GC_bis}  can
be generalized to ${\mathcal{T}}_{x\rightarrow y|z}$. In the linear
case, \citet{barnett2009granger} proved that ${G}_{x\rightarrow y|z}=2{\mathcal{T}}_{x\rightarrow y|z}$,
so that Granger causality through MAR(1) modelling is a special case
of causality defined through transfer entropy.

In general, any method which evaluates whether adding a new time series
$\mathbf{x}$ to a dynamical system for variables $y_{1},...,y_{n}$
improves prediction of $y_{i}$ can be defined as a generalised conditional GC method evaluating $x\rightarrow y_{i}|(y_{1},y_{2},...,y_{i-1},y_{i+1},...,y_{n}).$
Quite a number of nonlinear Granger causality inference techniques then fall within this category \citep[e.g.,][]{Marinazzo2008,Palus2008}.

\section{Additional results}

\subsection{Choice of p-values and thresholds on effect sizes \label{subsec:Choice-of-p-values}}

During preliminary simulations, we discovered that false causalities in absence of interactions could arise in larger proportion than the set false positive rate ($\alpha=$ 10\% for 2-species simulations, 20\% for 10- and 20-species simulations), probably due to inexact p-values. We thus searched for additional conditions on the estimates, such as effect sizes thresholds, to avoid relying on significance only. We based our analyses on the stochastic model described
in eqs.~\ref{eq:twoSpeRickerStoch_compet}--\ref{eq:twoSpeRickerStoch_compet_2} in the main text.

\begin{figure}[H]
\begin{centering}
\includegraphics[width=0.85\textwidth]{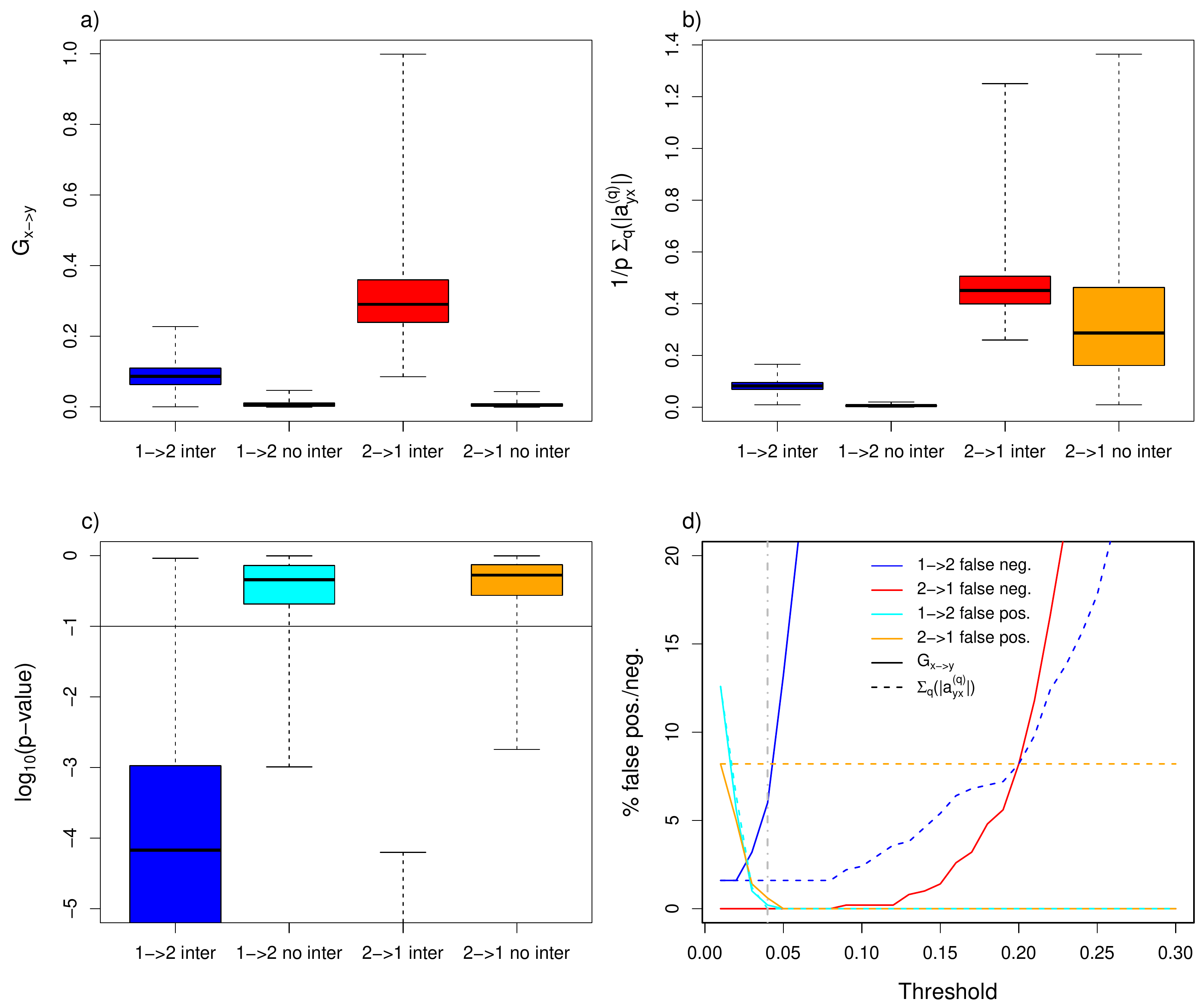}
\par\end{centering}
\caption{Comparison of methods to determine Granger-causality between two variables
in a stochastic model. Log ratio of residuals sum of squares (a) and
average effect of the causal species (b) are compared. The proportions
of false negatives (blue and red) and false positives (cyan and
orange), depending on the p-value and threshold (grey vertical line) imposed on these effects,
are shown in d) \label{fig:Comparison-of-methods_GC}}
\end{figure}

For Granger-causality, in order to consider effect sizes, we computed the log-ratio of the residuals
sum of squares (using notations from eqs. ~\ref{eq:directGC} and \ref{eq:GC_bis},
$\log\left(\frac{\sum\eta_{i}{{}^2}}{\sum\epsilon_{i}{{}^2}}\right)$
) as well as the average effect of the causal species over all causal lags up to $p$ ($\frac{\sum_{j}|a_{ij}|}{p}$).
We see on Fig. \ref{fig:Comparison-of-methods_GC} that the log-ratio
tends to be a more efficient indicator of causality and that fixing
a threshold of 0.04 for this log-ratio seems to achieve a good balance
between false negatives and positives.

\begin{figure}[H]
\begin{centering}
\includegraphics[width=0.85\textwidth]{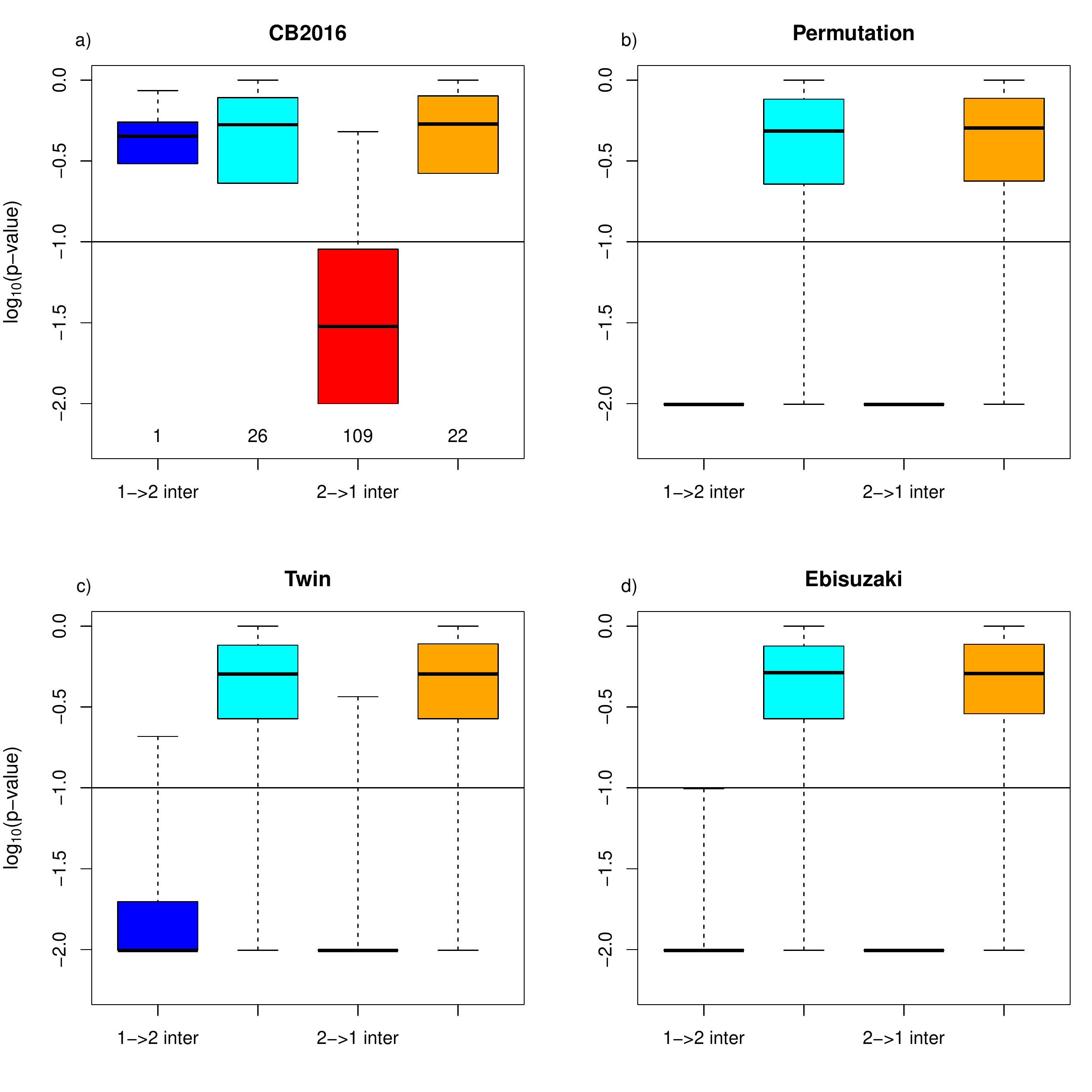}
\par\end{centering}
\caption{Log10(p-values) for the stochastic
2-species model, using different methods to compute p-values. \cite{cobey2016limits} method and permutation-, twin- and Ebisuzaki-based surrogates are  compared. Lighter colors (cyan and orange) represent simulations where there is no interaction, significant p-values are therefore indicators of false positives for these colors. The number of
p-values which are found to be 0, among the 500 simulations estimated,
is written at the bottom of the p-value boxplot. \label{fig:Comparison-pval-CCM}}
\end{figure}

\begin{figure}[H]
\begin{centering}
\includegraphics[width=0.85\textwidth]{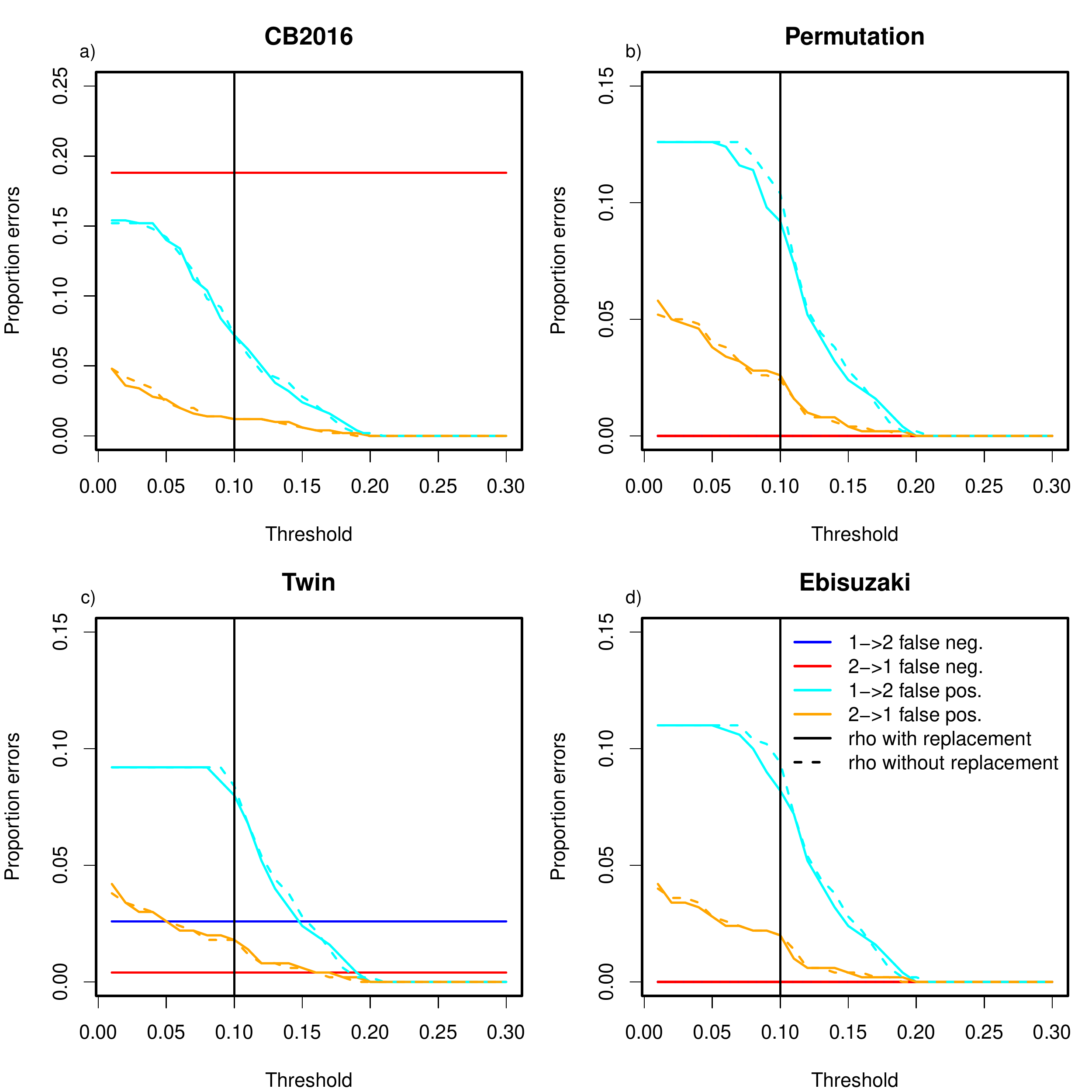}
\par\end{centering}
\caption{Comparison of the proportion of false negatives and false positives for CCM when combining p-values thresholds with thresholds on the cross-map skill $\rho$.\label{fig:Comparison-of-false-neg-pvalues}}
\end{figure}

For convergent cross mapping, how to compute the p-value itself
was a non-trivial issue (as discussed in Methods). We compared the p-value described
by \citet{cobey2016limits}, and three different types of surrogates implemented in \citet{ye2018redm}:
permutation, distance-based (`twin', the sampling replaces one point
by another which remains close in value) or frequency-based (`Ebisuzaki',
the time series spectrum is kept during resampling, \citealp{ebisuzaki_method_1997}). We also examined
the effect of putting a threshold on the value of $\rho$. We see
on Fig. \ref{fig:Comparison-pval-CCM} that surrogate-based p-values
are more efficient to detect causalities (and currently recommended by the rEDM team). As surrogate-based p-values have very similar
behaviors, we chose to keep the simplest (and least intensive computationally) method, based on permutation. That said, it should be kept in mind that for population dynamics with slower dynamics than considered here, reddening the frequency spectrum, other surrogates might perform best. We also considered an effect size threshold on $\rho$ values to avoid the majority of false positives and false negatives, values 0.1 or 0.2 were found efficient (Fig.~\ref{fig:Comparison-of-false-neg-pvalues}) and used when considering such thresholds in the main text tables.

\subsection{Effect of log-transformation on CCM}
\label{sec:log-transfo}
\begin{figure}[H]
\begin{centering}
\includegraphics[width=0.95\textwidth]{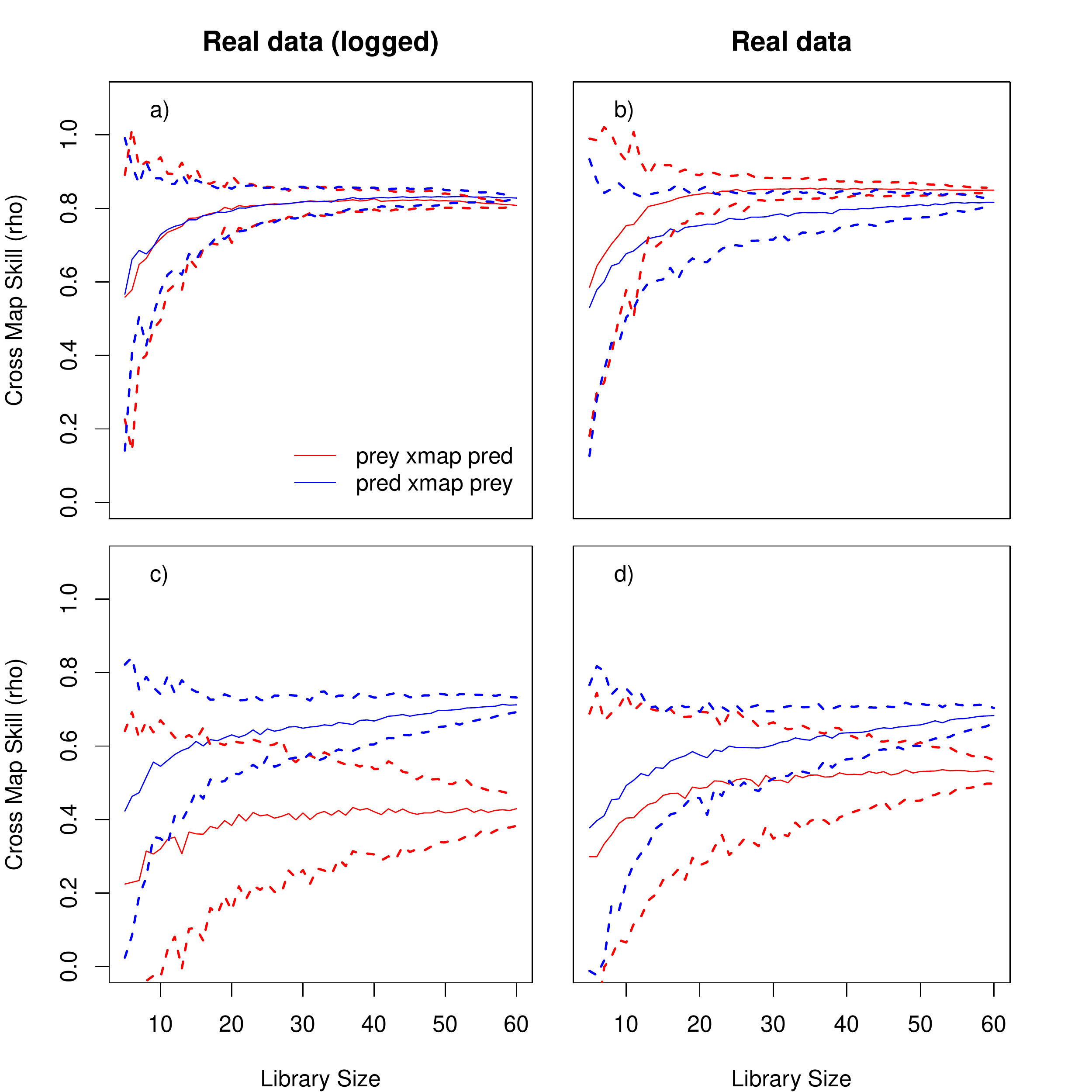}
\par\end{centering}
\caption{Convergent cross mapping with (left) and without (right) log-transformation
of the data for the Veilleux dataset. Top, CC0.5 dataset and bottom, CC0.375 dataset. \label{fig:log-transfo-veilleux}}
\end{figure}

\subsection{Deterministic two-species competition model}

\paragraph{Lag order $p$ selection in the MAR($p$) framework}

\begin{figure}[H]
\begin{centering}
\includegraphics[width=0.8\textwidth]{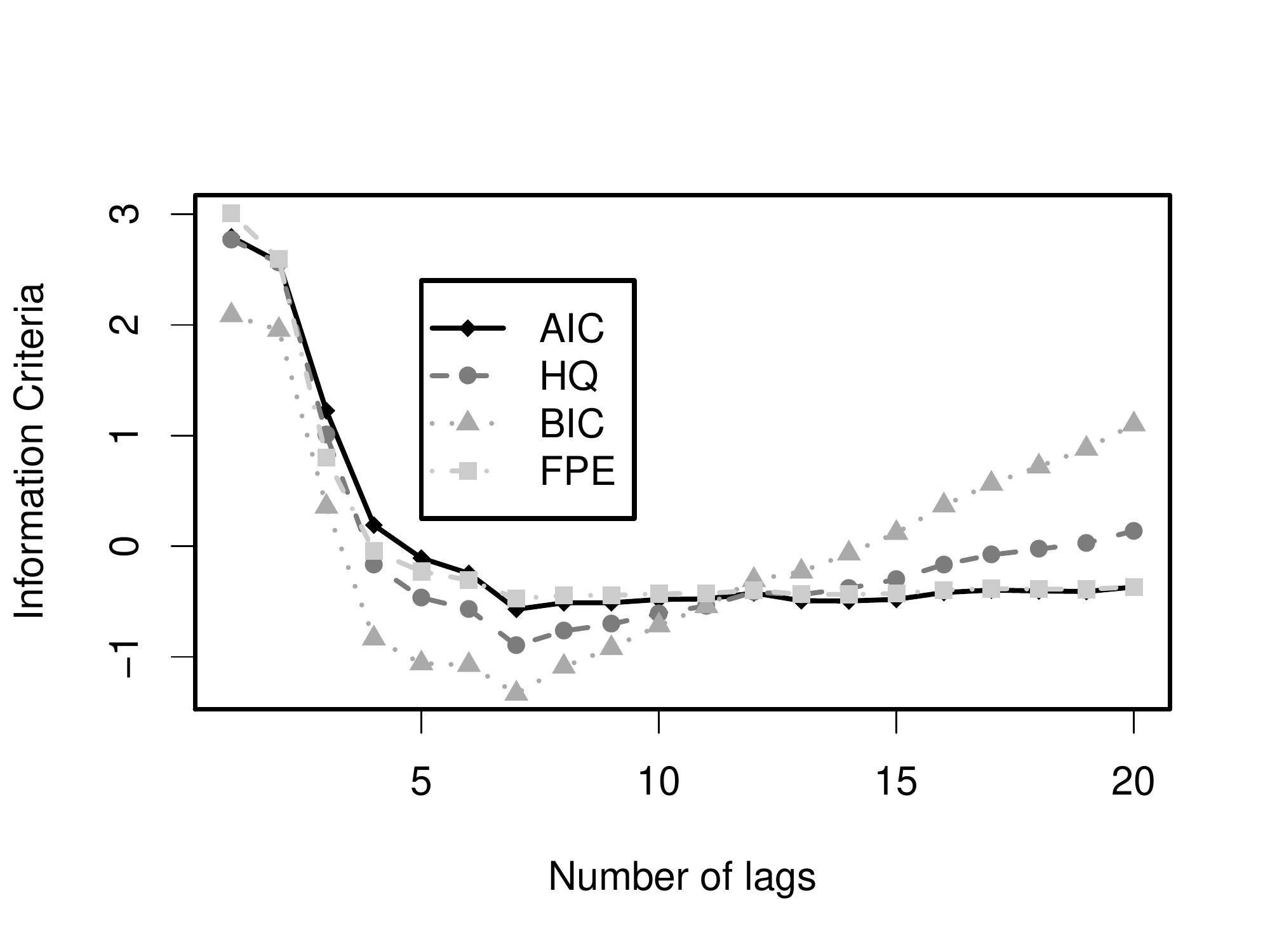}
\par\end{centering}
\caption{Model information criteria of MAR($p$) models,  as a function of lag order $p$, using the simulated deterministic competition model of eq. \ref{eq:two_species_chaotic_compet} in the main text as data.  \label{fig:ResultsDeterCompet_lagOrder}}
\end{figure}

\paragraph{Granger causality at various lag orders $p$}

\begin{figure}[H]
\begin{centering}
\includegraphics[width=0.95\textwidth]{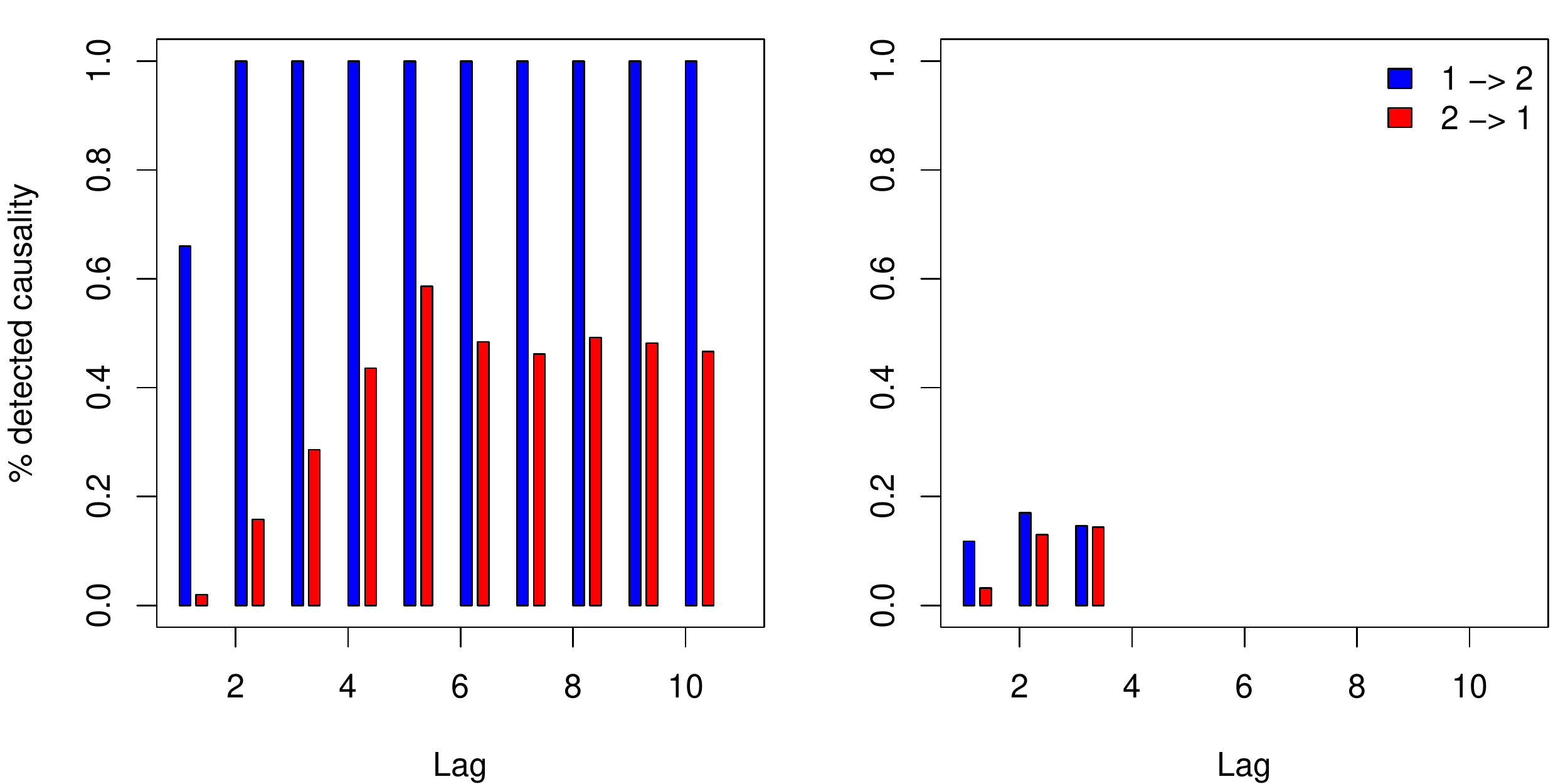}
\par\end{centering}
\caption{Proportion of detected Granger-causality, at the 10\% significance threshold, over 500 chaotic simulations with (left) and without (right) actual interactions between species, depending on the number of time lags taken into account (x-axis). Without interactions, the optimal lag is 3 and the Wald test cannot be performed for $p>3$.\label{fig:GC_deterministic}}
\end{figure}

\paragraph{Convergent cross mapping for 500 randomly drawn initial conditions}

\begin{figure}[H]
\begin{centering}
\includegraphics[width=0.95\textwidth]{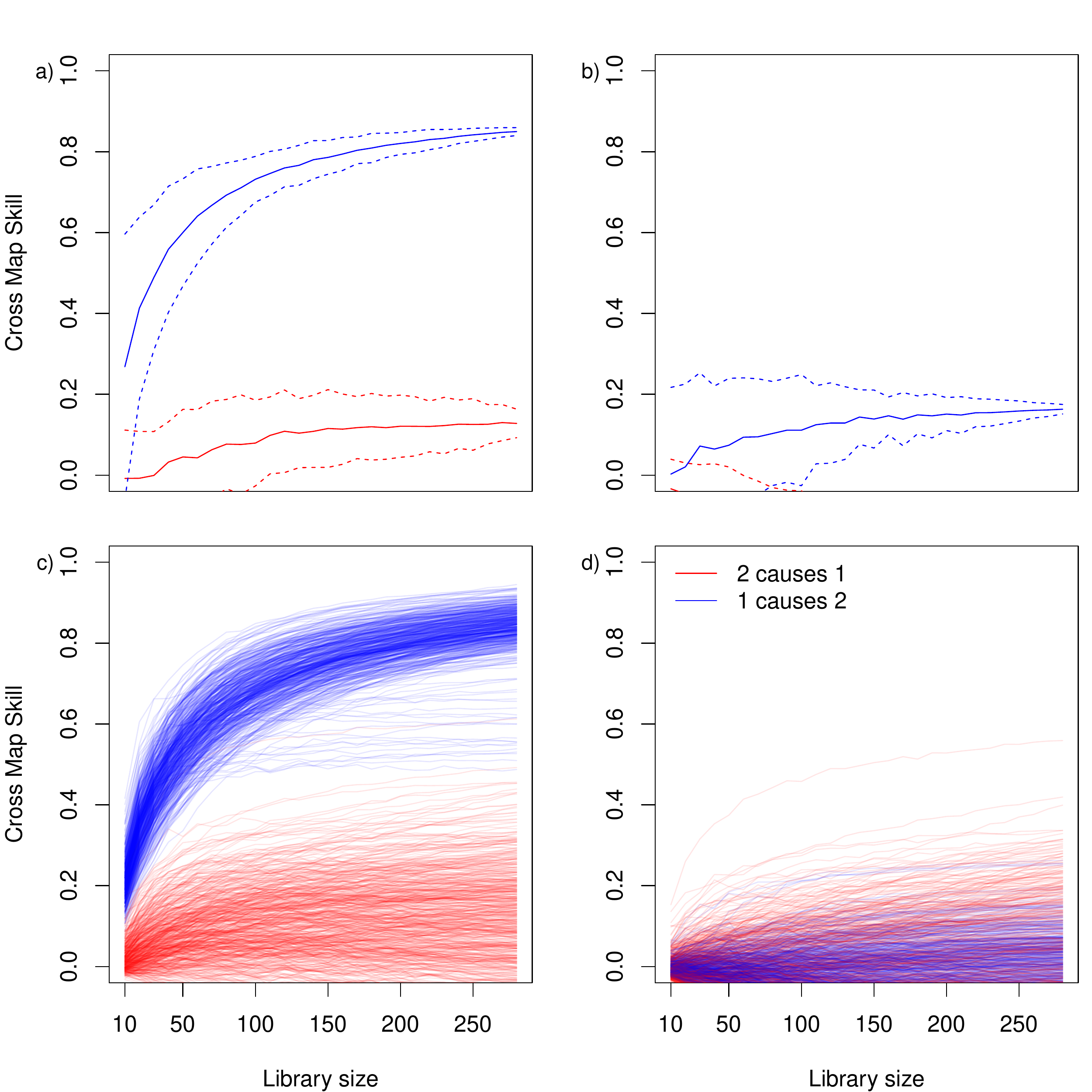}
\par\end{centering}
\caption{Convergent cross mapping on a simulated deterministic 2-species model, with (left) and without (right) competition between the two species. On the top
row, one simulation with (a) and without (b) interactions with associated
confidence bands (+/- 2 SD); bottom row, cross-map skill ($\rho$) for 500 simulations.\label{fig:CCM_Deter}}
\end{figure}




\subsection{GC applied to stochastic competition with a shared abiotic driver}\label{sec:GC-2spdriver}

\begin{figure}[H]
\begin{centering}
\includegraphics[width=1\textwidth]{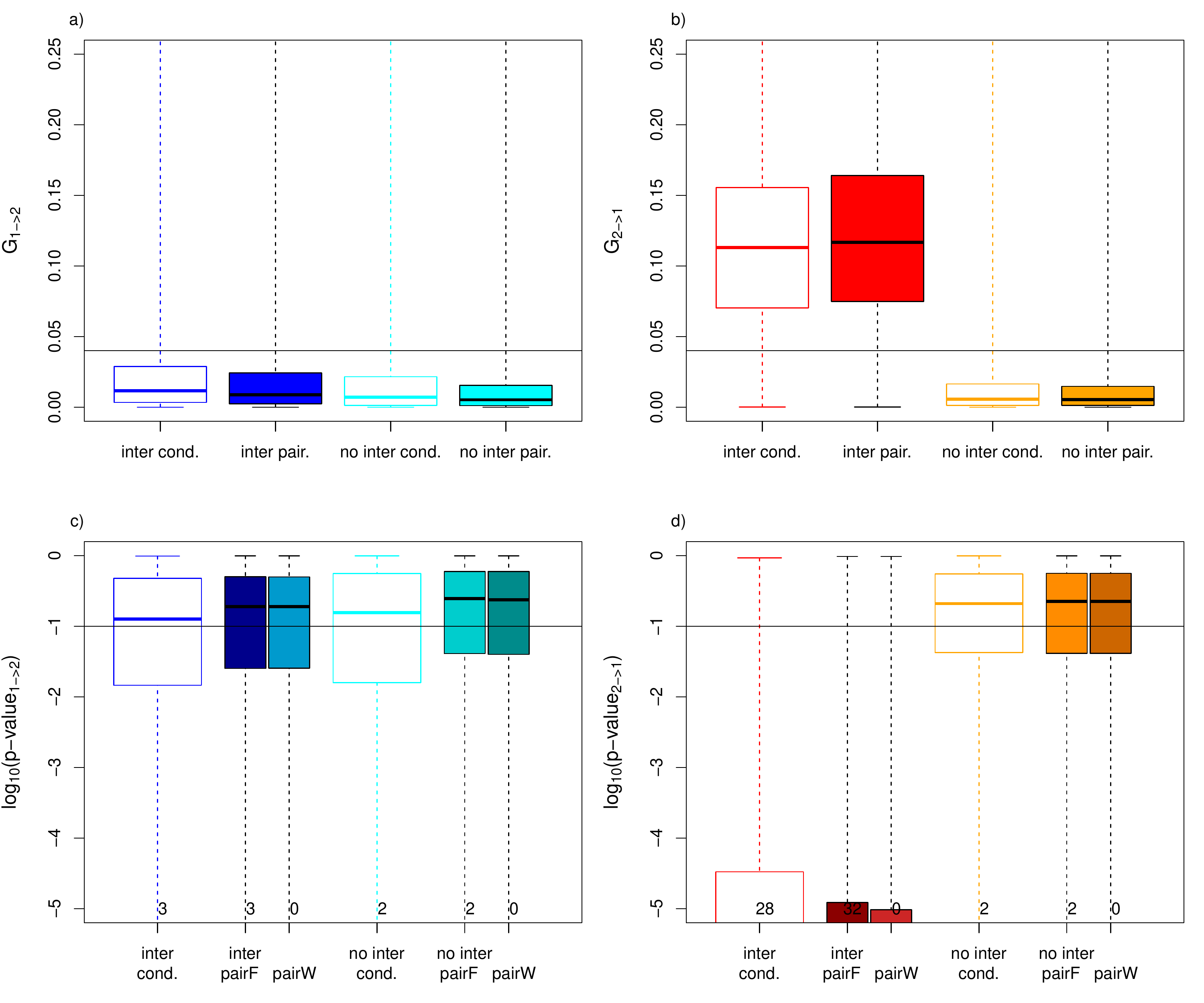}
\par\end{centering}
\caption{Log-ratio of residuals and log10(p-value) associated with the interactions from species 1 to species 2 (a and c) and from species 2 to species 1 (b and d), for a Granger-Causality analysis on 500 simulations of two species forced by an abiotic driver. Unfilled boxes represent interactions taking into account the driver (conditional) while filled boxes represent pairwise analyses. For pairwise analyses, two statistical tests were compared : F-test and Wald test. The number of p-values which are at 0 is indicated at the bottom of the p-value boxplots. Horizontal lines in a) and b) are the thresholds imposed to the effect size for the interaction to be considered relevant, while they correspond to the 10\% $\alpha$-threshold in c) and d). Blue color (resp. turquoise) refers to species 1 effects whenever there is a true effect (resp. when there is no true effect, `no inter'). Red color (resp. orange) refers to species 2 effects (yellow to brown when there is no true effect). The abbreviation `cond.' refers to conditional Granger causality while `pair' refers to pairwise Granger causality.  \label{fig:GC_driver} }

\end{figure}

\begin{figure}[H]
\protect\centering{}\protect\includegraphics[width=0.95\textwidth]{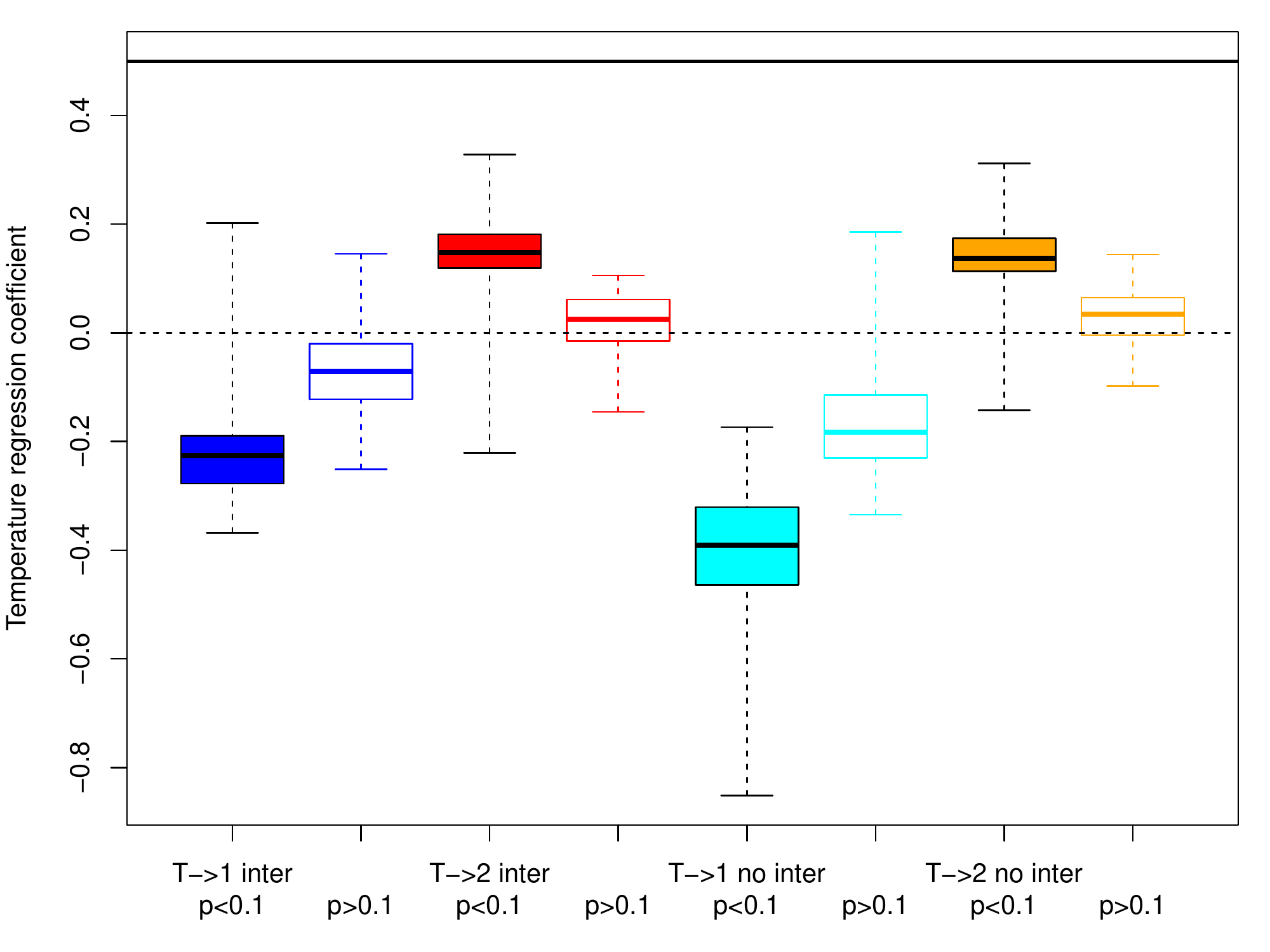}\protect\caption{Estimated effect of the abiotic driver on species 1 (blue, cyan) or species 2 (red, orange) when temperature is considered as an exogenous variable in the Granger-causality analysis. The value of the coefficient used in the simulations is indicated by the horizontal line at 0.5. Boxes with a coloured background correspond to a significant driver effect (at the 10\% level) while boxes with a white background correspond to effects which were deemed not significant. Light colours (cyan and orange) correspond to simulations in which there is no interaction between species.\label{fig:temp_effect_GC}}
\protect
\end{figure}

\subsection{CCM applied to stochastic competition with a shared abiotic driver}\label{CCm-2spdriver}

\begin{figure}[H]
\begin{centering}
\includegraphics[width=1\textwidth]{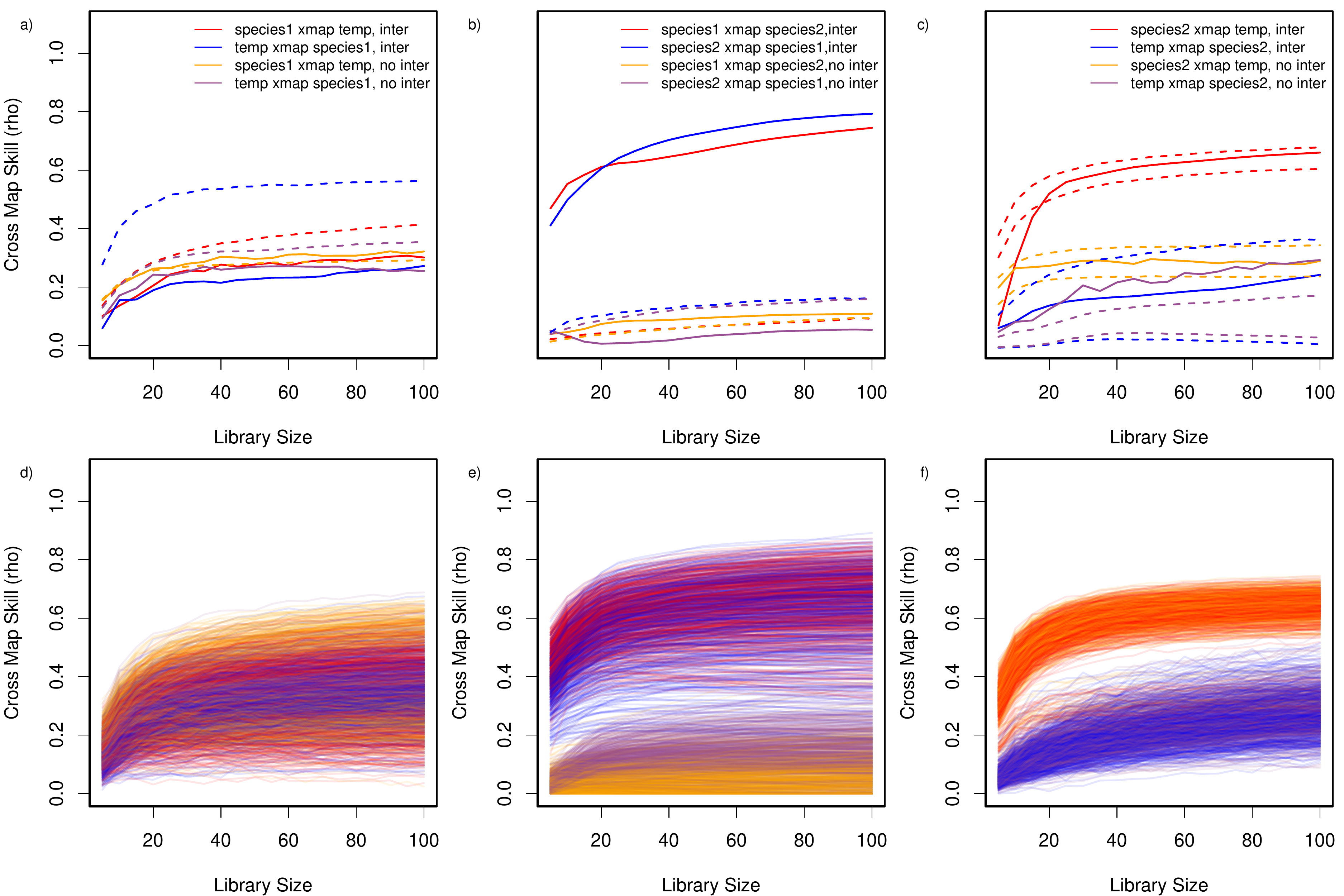}
\par\end{centering}
\caption{Convergent cross mapping for the two species forced by an environmental
driver (denoted as temp), when interactions are present (blue, red)
and when interactions are absent (purple, orange), for 500 simulations.
Dashed lines indicate the 10\% interval for rho-values obtained from
surrogate time series, i.e., time series that have the same seasonal
forcing but whose cross-correlations are altered.\label{fig:CCM_2sp_wDriver} }

\end{figure}

\subsection{Causality with respect to the abiotic driver}\label{sec:CCM-temperature}

\begin{figure}[H]
\protect\centering{}\protect\includegraphics[width=0.95\textwidth]{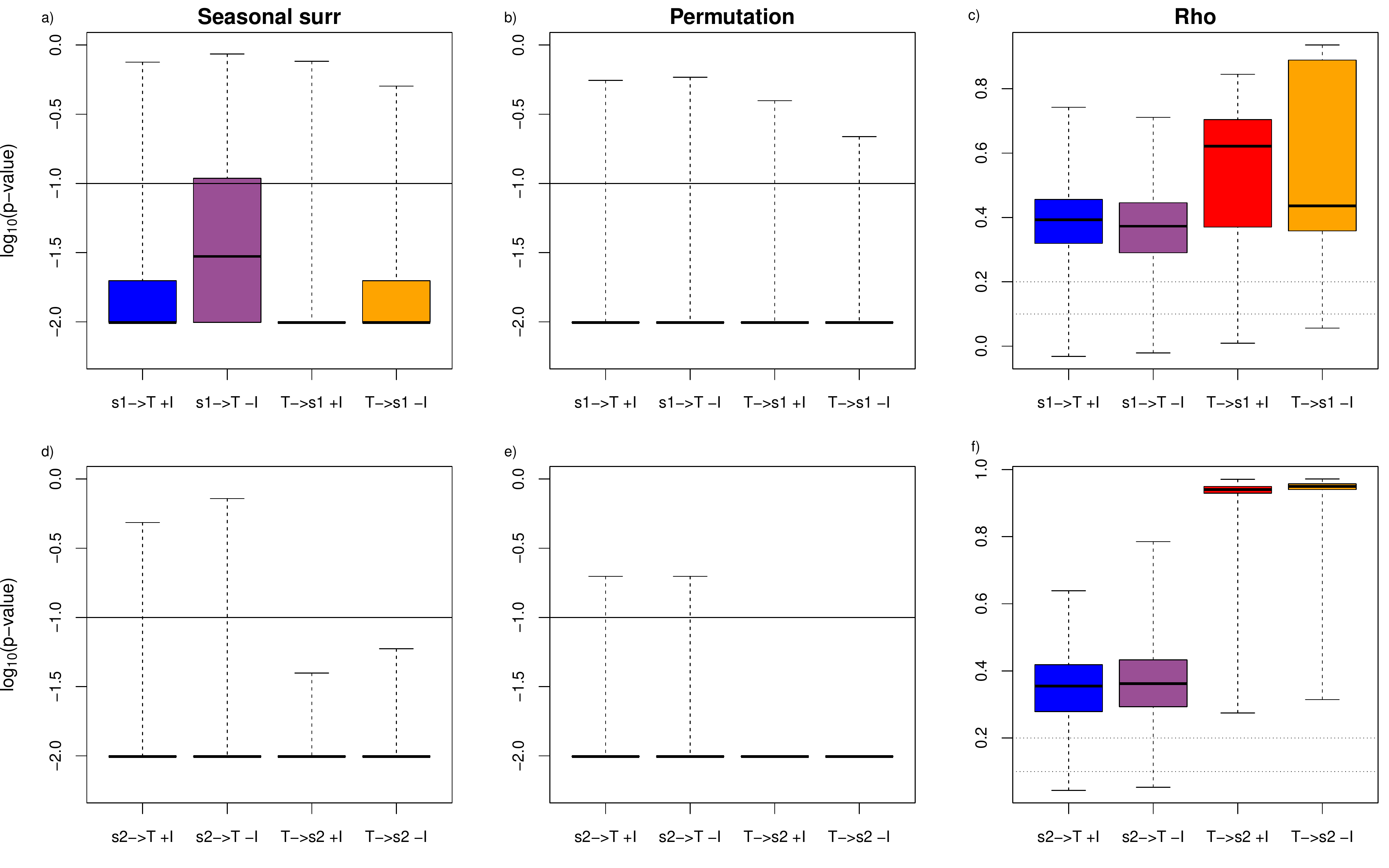}\protect\caption{Comparison of log10(p-values) and CCM skill ($\rho$) values to examine
effects of temperature (T) on species 1 and 2 (s1 and s2), and the
spurious reverse causality (species 1 or 2 causing temperature). Simulations
were ran with (+I) and without (-I) interactions between species 1
and 2. The 10\% false positive threshold
is indicated by a line on the pval plots (p-value must be below this
line for the causality to be inferred) while the 0.1 and 0.2 thresholds
that could be imposed on rho values are dotted lines in the right
panel ($\rho$  must be above the line for the causality to be inferred)\label{fig:Comparison-dummy-CCM}}
\protect
\end{figure}

\subsection{Lag order selection for the 10- and 20-species model}\label{sec:lagorder-HDsystems}

\begin{figure}[H]
\begin{centering}
\includegraphics[width=0.45\textwidth]{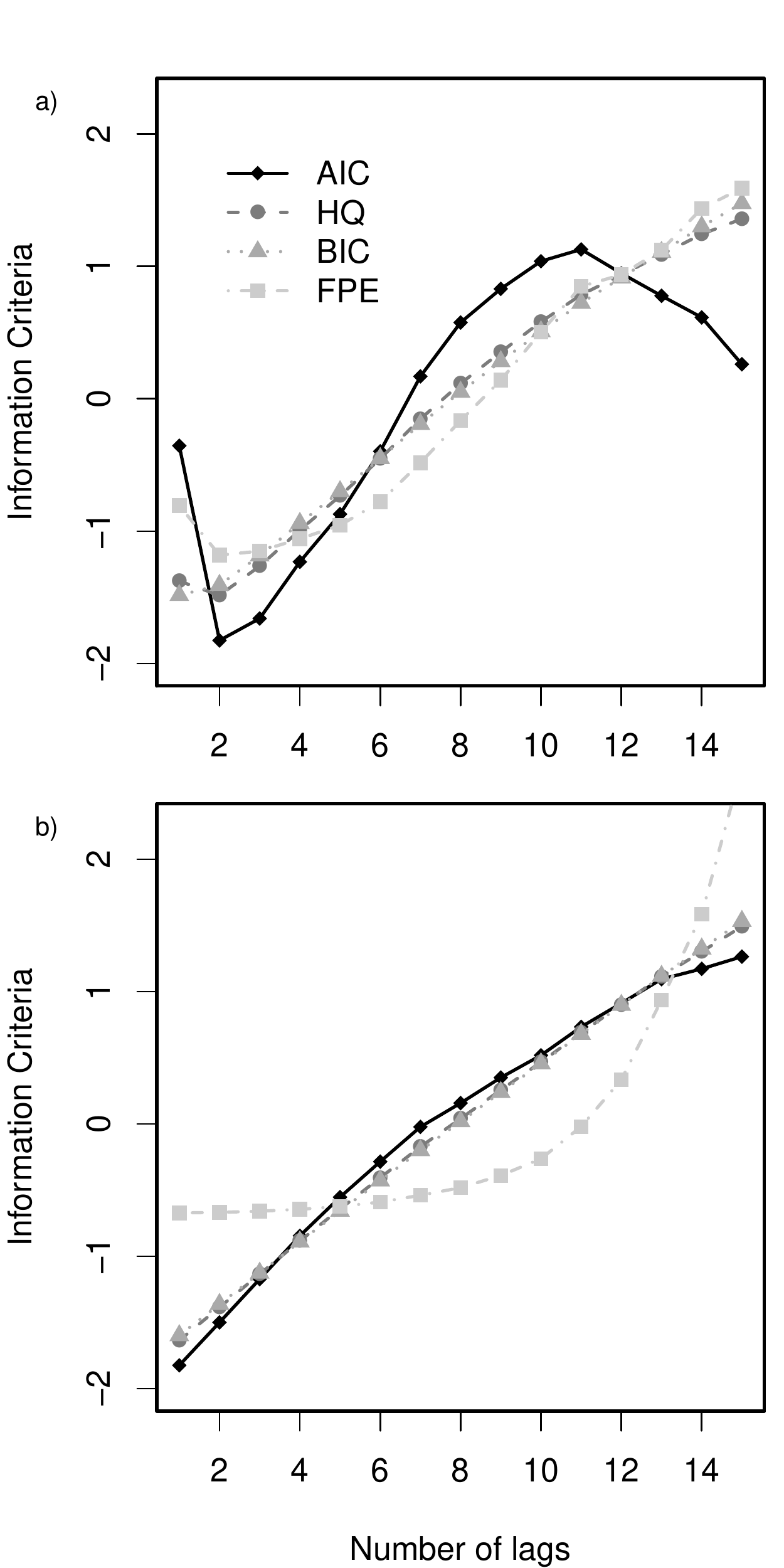}
\par\end{centering}
\caption{Lag order selection for (a) the 10-species and (b) one of the 20-species
stochastic community model.\textbf{\label{fig:Lag-order-selection_large-models-2}}}
\end{figure}

\subsection{Interaction matrix for the 20-species model}

\begin{figure}[H]
\includegraphics[width=0.95\textwidth]{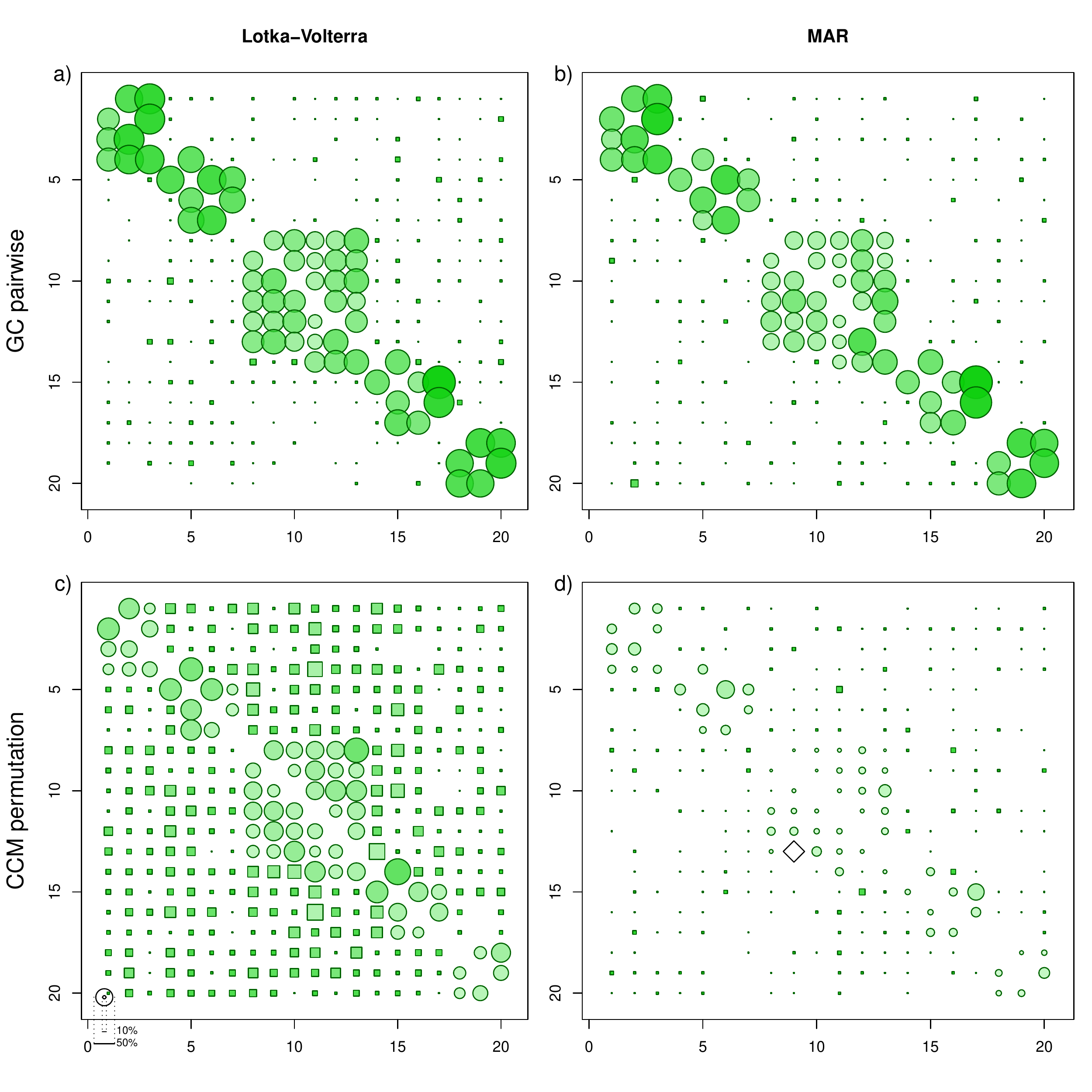}
\caption{Interaction matrices obtained from pairwise GC (top) or permutation-based surrogates for
CCM (bottom) for 20-species communities. Green circles represent true positives, green squares are false positives and empty diamonds are false
negatives. For true and false positives, the size of the symbols is proportional to the proportion of detection over 25 simulations. A symbol filled with a darker green represents a better performance (which happens with large circles or small squares).  \label{fig:Interaction-mat-20species}}
\end{figure}

\subsection{Alternative network reconstruction methods}
Here we present the interaction matrices for 10 and 20 species for the structured LASSO (GC) and Cobey-Baskerville p-values (CCM), which performed less well than the methods presented in the main text. 
\begin{figure}[H]
\includegraphics[width=0.95\textwidth]{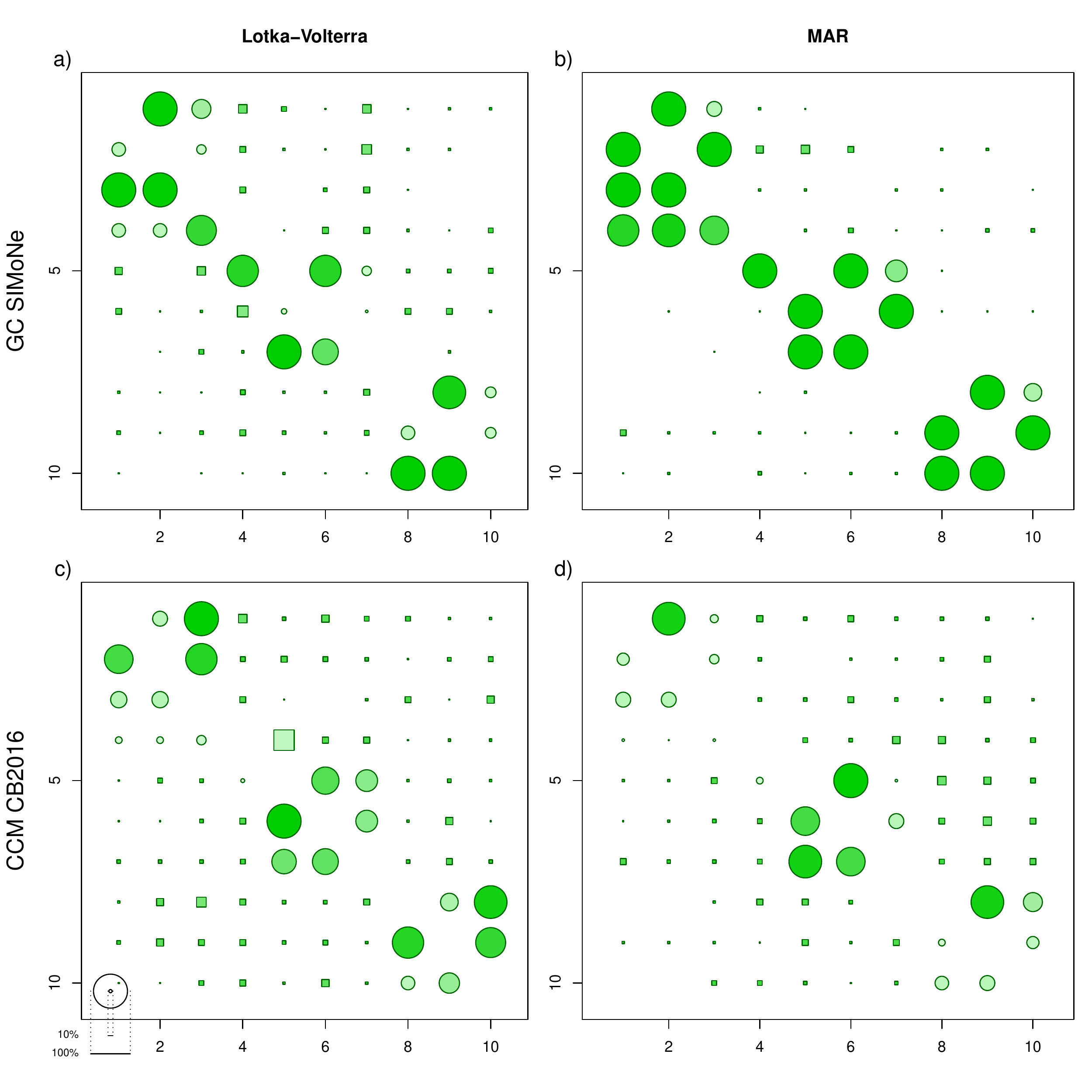}
\caption{Interaction matrices obtained from GC (top) or
CCM (bottom) for 10-species communities, based on alternative ways of computing p-values (see main text). Green circles represent true positives, green squares are false positives and empty diamonds are false
negatives. For true and false positives, the size of the symbols is proportional to the proportion of detection over 25 simulations. A symbol filled with a darker green represents a better performance (which happens with large circles or small squares). \label{fig:Interaction-mat-20species-bis}}
\end{figure}

\begin{figure}[H]
\includegraphics[width=0.95\textwidth]{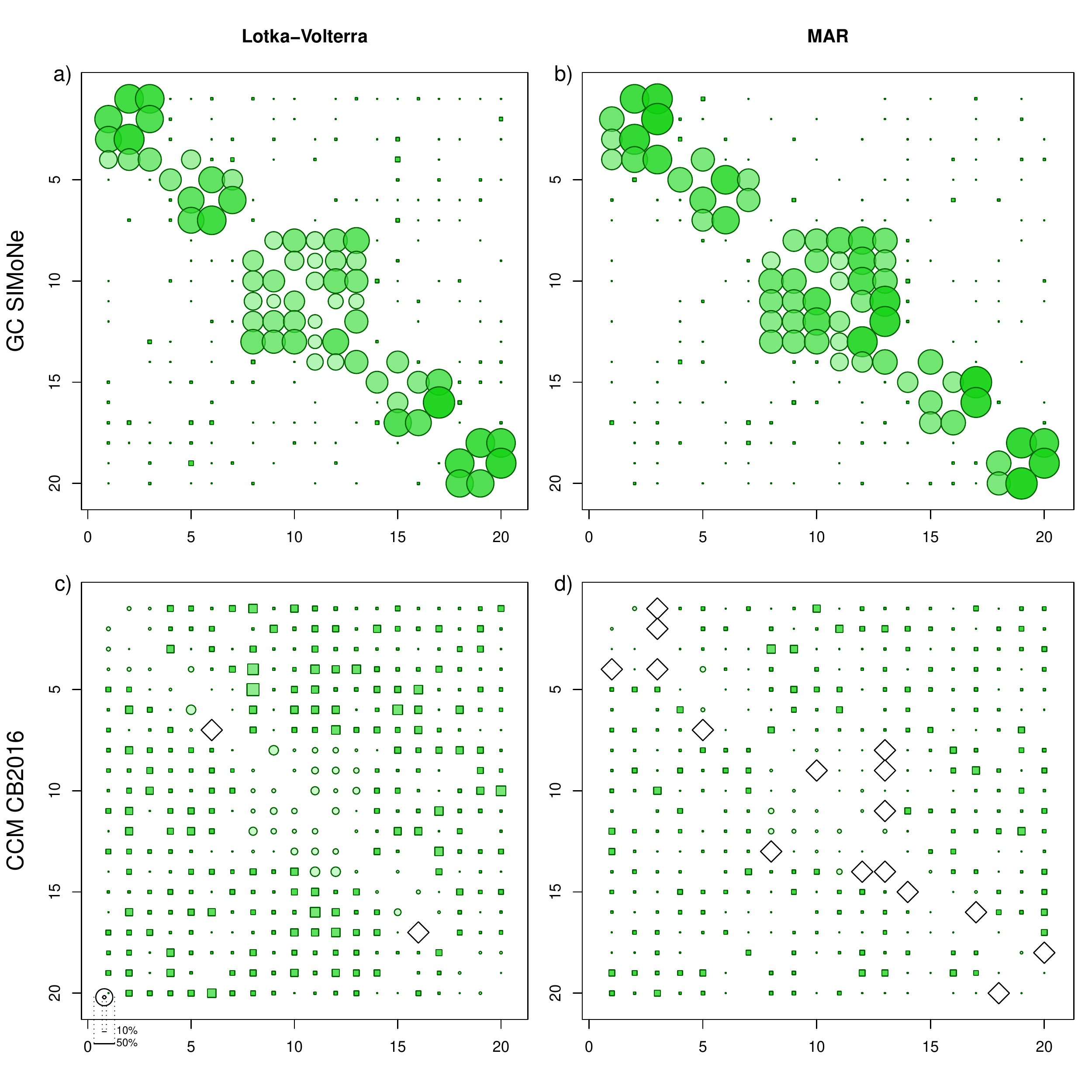}
\caption{Interaction matrices obtained from GC (top) or
CCM (bottom) for 20-species communities, based on alternative ways of computing p-values (see main text). Green circles represent true positives, green squares are false positives and empty diamonds are false
negatives. For true and false positives, the size of the symbols is proportional to the proportion of detection over 25 simulations. A symbol filled with a darker green represents a better performance (which happens with large circles or small squares). \label{fig:Interaction-mat-20species-ter}}
\end{figure}

\end{document}